\documentclass{JHEP3}
\usepackage{amsmath,epsfig}

\newcommand{\be}{\begin{equation}}
\newcommand{\ee}{\end{equation}}
\newcommand{\bea}{\begin{eqnarray}}
\newcommand{\eea}{\end{eqnarray}}

\newcommand{\ba}{\begin{array}}
\newcommand{\ea}{\end{array}}

\newcommand{\Dslash}{\relax{\kern+.25em / \kern-.70em D}}

\newcommand{\Real}{\relax{\mathsf{\Gamma\kern-.35em R}}}
\newcommand{\Int}{\relax{\mathsf{Z\kern-.40em Z}}}

\newcommand{\gbar}{\kern1pt\overline{\kern-1pt g\kern-0pt}\kern1pt}
\newcommand{\mbar}{\kern2pt\overline{\kern-1pt m\kern-1pt}\kern1pt}
\newcommand{\obar}[1]{\kern3pt\overline{\kern-2pt #1\kern-0pt}\kern1pt}

\newcommand{\abar}{\kern1pt\overline{\kern-1pt a\kern-0.5pt}\kern1pt}

\title{Minimal models with light sterile neutrinos }
\preprint{
	IFIC/11-23\\
	IFT-UAM/CSIC-11-29\\
	PPP/11/23
         DCPT/11/46\\
         EURONU-WP6-11-36\\
	}

\author{A. Donini$^{1,2}$, P.~Hern\'andez$^1$, J.~L\'opez-Pav\'on$^3$ and M.~Maltoni$^2$\\
       (1)  Instituto de F\'{\i}sica Corpuscular, CSIC-Universitat de Val\`encia\\
		Apartado de Correos 22085, E-46071 Valencia, Spain\\
		(2)Instituto de F\'{\i}sica Te\'orica UAM/CSIC, Calle Nicol\'as Cabrera 13-15, E-28049 Madrid, Spain \\
		(3)Institute for Particle Physics Phenomenology (IPPP), Department of Physics, Durham
University, Durham DH1 3LE, UK
 \\
}

\abstract{
We study the constraints imposed by neutrino oscillation experiments on the minimal extensions of the Standard Model (SM) with  $n_R$ gauge  singlet fermions (``right-handed neutrinos''), that can account for neutrino masses. We consider the most general coupling  of the new fields to the SM fields, in particular those that break lepton number and we do not assume any a priori hierarchy in the mass parameters. We proceed to analyze these models starting from the lowest level of complexity, defined by the number of extra fermionic degrees of freedom. The simplest choice that has enough free parameters in principle (i.e. two mass differences and two angles) to explain the confirmed solar and atmospheric oscillations corresponds to $n_R=1$. This minimal choice is  shown to be excluded by data. The next-to-minimal choice corresponds to $n_R=2$. We perform a systematic study of the full parameter space in the limit 
of degenerate Majorana masses  by requiring that at least two neutrino mass differences correspond to those established by solar and atmospheric oscillations. We identify several types of  spectra that can fit long-baseline reactor and accelerator neutrino oscillation data, but fail in explaining solar and/or atmospheric data. The only two solutions that survive are the expected seesaw and quasi-Dirac regions, for which we set
lower and upper bounds respectively  on the Majorana mass scale. Solar data from neutral current measurements provide essential information to constrain the quasi-Dirac region. The possibility to accommodate the LSND/MiniBoone and reactor anomalies, and the implications for neutrinoless double-beta decay and tritium beta decay are briefly discussed.   }

\begin{document}

\section{Introduction}

The existence of tiny neutrino masses is probably the first signal of 
physics beyond the Standard Model (SM). One can envision various possibilities for such new physics, but the simplest is probably 
the addition of extra singlet fermions, that can play the role of the missing helicity partners of the SM neutrinos. Such models include very different possibilities ranging from Dirac neutrinos to the popular Type-I seesaw models \cite{Minkowski:1977sc,GellMann:1980vs,Yanagida:1979as,Mohapatra:1979ia}. The particle content is the same in both cases, but the couplings of the extra especies to the SM fields and among themselves are different. 

In the case of seesaw models, the most general couplings are assumed, in particular those that break lepton number, such as Majorana masses for the singlet fermions. In order to recover naturally light neutrinos, those masses are assumed to be very large, above the electroweak scale. There is however no reason why this scale should be so high,  and several works have considered the phenomenological implications of a much lower seesaw scale \cite{deGouvea:2005er,deGouvea:2006gz,Liao:2006rn,Nelson:2010hz}. Dirac neutrinos can be recovered by imposing a global lepton number symmetry. Such a lepton number symmetry can however be implemented in more than one way, since there is  freedom  in the lepton number charge assignments of the extra sterile species. Different choices of these charges led to different models besides Dirac neutrinos, such as the inverse seesaw  \cite{Wyler:1982dd,Mohapatra:1986bd}.

The purpose of this paper is  a first step towards  a systematic exploration of the phenomenology of such models, in increasing order of complexity, in order to quantify the constraints imposed by data. The theoretical prejudice will be just the number of extra Weyl fermions, $n_R$. We consider therefore the more general couplings, i.e. without imposing lepton number symmetries, although obviously the models with exact symmetries can be recovered in appropriate limits.
In this work we will consider just the simplest possibilities $n_R=1,2$ and leave other cases for future investigations.

The simplest choice, which corresponds to adding just one additional singlet Weyl fermion, $n_R=1$, 
 has in principle  enough free parameters to fit the solar and atmospheric oscillation data. A more detailed analysis however shows that  it does not work, just considering long-baseline oscillation neutrino data. 
The next-to-minimal choice requires two Weyl fermions, $n_R=2$. Such a possibility is of course well known to be a viable one (as long as the LSND/MinBoone signal is discarded), both in the Dirac limit and in the standard seesaw limit. In both cases, the physics spectrum contains one massless neutrino and two massive ones and, therefore, the two required splittings. The main goal of the paper will be the exploration of the parameter space in between these two limits, to search for other viable solutions that could accommodate at least the solar and atmospheric oscillation. In order to analyse this model we discuss a convenient parametrization valid for $n_R \leq 3$. In the mini-seesaw region, the Casas-Ibarra parametrization \cite{Casas:2001sr} is used to derive approximate oscilllation probabilities, accurate in this regime.

Many works before have been devoted to study the implications of neutrino oscillation data on models with extra sterile neutrinos (some recent analyses are \cite{Sorel:2003hf,Akhmedov:2010vy,Kopp:2011qd,Giunti:2010jt}), usually refered to as $3+1, 3+2, ...3+N_s$. Most of these studies have been done with the motivation of trying to accommodate LSND \cite{Aguilar:2001ty}, and MiniBooNe data \cite{mb2,mb1}.  It is important to stress that these phenomenological models usually correspond to a generic model with $3+N_s$ mass eigenstates. The number of  free parameters for  $N_s=n_R$  is typically much larger  than what we find, either because the number of Weyl fermions involved is different (e.g. $3+1$ Dirac fermions correspond in our context to $n_R=5$ and not to $n_R=1$) or because couplings that are forbidden by gauge invariance in our model, such as Majorana mass entries for the active neutrinos, are included effectively in the phenomenological models. It is obvious that our models for any $n_R$ will be contained as restricted cases in these phenomenological models for $N_s=n_R$. However, the analyses performed in these works do not take such limits and usually restrict the 
number of parameters by assuming instead some hierarchies between neutrino masses that could accommodate LSND.

The structure of the paper is as follows. We start by setting up the notation in sec.~\ref{sec:models} and the parametrization in sec. \ref{sec:param}. In sec.~\ref{sec:3+1} we consider the implication of oscillation experiments on the 3+1 model. In sec.~\ref{sec:3+2} we consider the 3+2 model in the degenerate limit in the context of LBL, solar and atmospheric oscillations.  In sec.~\ref{sec:lsnd} we discuss the implications for LSND/MiniBoone, as well as for tritium and neutrinoless double beta decay experiments. In sec.~\ref{sec:nodeg} we briefly consider the non-degenerate case  in the mini-seesaw regime. The very stringent bound on the quasi-Dirac region implied by solar data is discussed in detail in sec.~\ref{sec:solar}. 

\section{Models with extra singlet fermions}
\label{sec:models}

The leptonic part of the Standard Model Lagrangian is: 
\be
{\cal L}_{SM} = \sum_{\alpha= 1}^3 \left \{ \bar l^\alpha_L \gamma^\mu D_\mu l^\alpha_L 
                                        + \bar e^\alpha_R \gamma^\mu D_\mu e^\alpha_R
                                        - \sum_{\beta = 1}^3 \bar l^\alpha_L Y^{\alpha \beta}_e \Phi e^\beta_R + h.c.,
\right \}
\ee
where $l^\alpha$ is the lepton doublet. To simplify notation, from now on summation over repeated indices is understood. 

We study the minimal extension of the Standard Model that can account for neutrino masses, i.e. the inclusion of 
$n_R$ gauge singlet Weyl fermions (hereafter called ``right-handed neutrinos'', $\nu_R$): 
\be
{\cal L} = {\cal L}_{SM} - \bar l^\alpha_L Y^{\alpha \beta}_\nu \tilde \Phi \nu^\beta_R + h.c.
\ee
If lepton number is violated, gauge-invariant Majorana mass terms can be added to the minimal
Lagrangian: 
\be
{\cal L} = {\cal L}_{SM} - \bar l^\alpha_L Y^{\alpha \beta}_\nu \tilde \Phi \nu^\beta_R 
        - \frac{1}{2} \bar \nu^{\alpha c}_R M^{\alpha \beta}_N \nu^\beta_R + h.c.
\ee

The  new mass parameters in the matrix $M^{\alpha\beta}_N$ are unbounded. If they are very large compared to the electro-weak symmetry breaking scale, $v$, the right-handed neutrinos decouple from the light spectrum. They can be integrated out giving mass 
to the light neutrinos through the effective dimension-5 operator \cite{Weinberg:1979sa}, 
\be
O_5 = \bar l^\alpha_L \left [Y_\nu M^{-1}_N Y^T_\nu \right]^{\alpha \beta} \tilde \Phi ~\Phi ~l_L^\beta.
\label{eq:weinbergop}
\ee
which induces, upon spontaneous symmetry breaking, a Majorana mass to the SM neutrinos of the form
\be
m_\nu = v^2 \left [ Y_\nu M^{-1}_N Y^T_\nu \right ]. 
\label{eq:numass}
\ee
This is the famous seesaw mechanism \cite{Minkowski:1977sc,GellMann:1980vs,Yanagida:1979as,Mohapatra:1979ia}, that can lead to small neutrino masses, even if their Yukawa couplings are of ${\mathcal O}(1)$, provided there is a large hierarchy between $v$ and $M_N$.

If the Yukawa coupling of $\nu_L$ and $\nu_R$ fermions with the Higgs field are, however, of the size of the electron 
Yukawa coupling or smaller,  $M_N = O(1)$ TeV or less and new interesting signatures are possible. 
 If $M_N = {\mathcal O}(100)$ GeV, 
  direct production of right-handed neutrinos through Higgs decay can take place at the LHC \cite{Bajc:2006ia,Garayoa:2007fw,Perez:2008ha,delAguila:2008cj}. For 
$M_N \in [0.1,10]$ GeV, right-handed Majorana neutrinos induce lepton-violating processes in meson decays (for a recent study see \cite{Atre:2009rg}). In the keV range
right-handed neutrinos can be interesting dark matter candidates (for a recent revival see \cite{Asaka:2006ek,Asaka:2006nq}) or in the eV range, they could contribute to explain the LSND-MiniBooNE anomaly \cite{Sorel:2003hf,Kopp:2011qd} and affect cosmological observables, such as the abundance of light elements, the anisotropies in the cosmic microwave background or the large scale structure (for recent analyses see \cite{GonzalezGarcia:2010un,Hamann:2010bk,Giusarma:2011ex}). 

In this paper we concentrate only on the constraints imposed by oscillation data which can be relevant for $M_N$ not much larger than 
a few eV.

\subsection{Physical parameters}

The number of physical parameters can be determined in different ways. In the following we use the method proposed in \cite{Santamaria:1993ah}. If $R/I$ is the number of real/imaginary parameters in the generic Yukawa matrices, $R^s/I^s$ is the number of real/imaginary parameters defining the symmetry group elements of the lepton sector when the Yukawa matrices are switched off, and $R^r/I^r$ is the number of parameters describing the 
group elements that survive for non-zero Yukawa matrices, the number of real/imaginary physical parameters should satisfy the following relation:
\begin{eqnarray}
R_{phys} &=& R - (R^s - R^r) , \nonumber\\
I_{phys} &=& I -(I^s - I^r) .
\end{eqnarray}
We have two separate equations for the real and imaginary parameters: the first ones become either masses or angles and the latter become phases. 

Let us do the counting in general for $n_L$ families of active and $n_R$ families of steriles. The charged lepton Yukawa matrix is $n_L \times n_L$, while the charged-neutral one is rectangular $n_L \times n_R$. The mass matrix of the Majorana neutrinos is a symmetric square matrix $n_R \times n_R$. The total number of real and imaginary parameters is, therefore:
\begin{eqnarray}
R = I = n_L (n_L + n_R) + n_R {(n_R+1)\over 2}.
\end{eqnarray}
If  all those matrices would vanish, the symmetry group of the lepton sector would be $U(n_L)$ for the left doublets, $U(n_R)$ for the Majorana neutrinos and 
$U(n_L)$ for the right-handed charged fermions. Therefore the number of real and imaginary parameters of the corresponding group elements is:
\begin{eqnarray}
R^s &=& {n_L (n_L-1)} + {n_R (n_R-1) \over 2}, \nonumber\\
I^s &=&  {n_L (n_L+1)} + {n_R (n_R+1) \over 2}.
\end{eqnarray}
There is no remaining symmetry when the masses are non-zero $R^r, I^r=0$, and therefore the number of physical parameters is:
\begin{eqnarray}
R_{phys} &=& n_L (n_R +1) + n_R = n_L n_R + n_R + n_L, \nonumber\\
I_{phys} &=&  {n_L (n_R-1)}.
\end{eqnarray}
Since the number of zero modes in the mass matrix is generically $n_L - n_R$, the number of mass parameters is  $n_L + n_R +  n_R = n_L + 2 n_R$ for $n_L \geq n_R$ (which should be real parameters). Therefore:
\begin{eqnarray}
N_{angles} &=& n_R (n_L -1)  \;\;\; n_L \geq n_R \nonumber\\
N_{phases} &=&  {n_L (n_R-1)} .
\end{eqnarray}
It is important to stress here again the difference between the model we are considering and the phenomenological $3+N_s$ models. For example the popular $3+1$ model has six physical angles and four mass eigenstates. In order to have that number of physical angles in our case we need to have at least $n_R = 3$, but that choice  would generically imply six mass eigenstates. In order to recover only four we would need to have some degeneracies, that is, some Weyl fermions must pair up into Dirac neutrinos, which typically reduces also the number of physical angles. For example, we could get the $3+1$ model from $n_R= 5$ and imposing a global lepton number symmetry (there are probably  other choices to obtain the same result).

A more systematic way to constrain models with singlet fermions is to classify them in increasing order of complexity, according to  the number of extra fermionic field degrees of freedom and the global symmetries, in contrast with the classification based on the physical spectrum that might represent different choices of the former type. In this work we will concentrate on the two simplest cases: $n_R=1$ and 2. The number of physical parameters and the generic spectrum for the different choices of global symmetries are summarized in Table~\ref{tab:param}.

\begin{table}
\begin{center}
\begin{tabular}{|l|l|l|l|l|l|}
\hline
$n_R$ & $L_i$ & $\#$ zero modes & $\#$ masses & $\#$ angles & $\#$ CP phases \\
\hline
 1 & - & 2 & 2 & 2 & 0 \\
  & +1 & 2 & 1 & 2 & 0 \\
 \hline
 2 & - & 1 & 4 & 4 & 3 \\
  & (+1,+1) & 1 & 2 & 3 & 1 \\
  & (+1,-1) & 3 & 1 & 3  & 1  \\
  \hline
  3 & - & 0 & 6 & 6 & 6 \\
  & (+1,+1,+1) & 0 & 3 & 3 & 1 \\
  & (+1,-1,+1) & 2 & 2 & 6  & 4
    \\
  & (+1,-1,-1) & 4 & 1 & 4   & 1  \\
  \hline
\end{tabular}
\label{tab:param}
\caption{Spectrum and number of independent angles and phases for the models with $n_R=1, 2$ without and with  global lepton number symmetries. The second column shows the lepton number, $L$, charge assignments of the extra singlets., $L_i$ Only charge assignments were none of the extra singlets gets completely decoupled are considered.}
\end{center}
\end{table}
In the absence of global lepton number symmetries, the case $n_R=1$ gives a spectrum of two massive and two massless fermions. There are two physical angles and no CP violation. If a lepton number symmetry is imposed,  there is only one charge assignment for the extra field that allows a renormalizable coupling to the SM neutrino fields  ($L=1$) \footnote{Any charge assignment that forbids couplings of the singlet fermions to the SM fields is of course uninteresting. }. In this case, the two massive Majorana fermions are degenerate and form a  massive Dirac fermion, while two Weyl fermions remain massless. 
Clearly  the model with $n_R=1$ with a global lepton number symmetry is ruled out, as it cannot accommodate the solar and atmospheric oscillations. On the other hand the model without the global symmetry contains in principle sufficient parameters (two mass eigenstates and two angles) to explain both oscillations lengths. 

The model with $n_R=2$ and no lepton number symmetries, gives rise to a spectrum including  four massive  and one massless neutrino.  There are also four physical angles and three CP violating phases. 

Simplifications also occur when lepton number symmetries are imposed. For $n_R=2$  there are two choices for the lepton number charge assignments that allow renormalizable couplings between the extra singlet fermions and the SM neutrinos. One obvious choice is to give both of the sterile fields lepton number charge +1. In this case, the spectrum degenerates into a massless neutrino and two massive Dirac neutrinos. The number of physical angles gets reduced to three and there is only one physical CP phase. Obviously this choice is as good as the standard three-neutrino mixing model to accommodate existing oscillation data. 

The other choice for the charge assignments is to give charge +1 only to one of the extra fields and -1 to the other. In this case, the spectrum consist of three massless neutrinos and one massive Dirac one. The number of physical angles is reduced to two and there is no CP violation. This model with just one mass cannot explain oscillation data. However, a small perturbation that breaks the lepton number symmetry is again as rich as  the generic case of $n_R=2$, but  with some strong hierarchies, naturally preserved by the approximate lepton number symmetry. This is the minimal flavour violating seesaw model considered in \cite{Gavela:2009cd} (see also \cite{Alonso:2010wu}). For $n_R=3$, there are many more possibilities, listed in Table~\ref{tab:param}.

\section{Parametrization}
\label{sec:param}

Consider the generic mass matrix corresponding to the model with three left-handed neutrinos and $n_R$ 
right-handed Majorana neutrinos: 
\begin{equation}
\label{massmatrix}
{\cal M_\nu} = \left ( \begin{array}{cc}
 0                & Y_\nu {v\over \sqrt{2}}    \\
 Y^T_\nu   {v\over\sqrt{2}}            & M_N   \\
                \end{array} \right )
\end{equation}
where $M_N$ is a $n_R\times n_R$ matrix and $Y_\nu$ is a $3\times n_R$ matrix. On the other hand,
without loss of generality, the $M_N$ matrix can be chosen as a diagonal matrix. For simplicity we will restrict to the case where all entries are real, so that there is no CP violation. 

The first step will be to do a rotation to bring the block $Y_\nu v/\sqrt{2}$ to a {\it minimal}  form where the rows $1,..,3-n_R$ can be set to zero and  the remaining squared  block is off-diagonal
\begin{equation}
\label{massmatrix}
{\cal M_\nu}' = \Omega^T    {\cal M_\nu} \Omega  =  \left ( \begin{array}{ccc}
0            & m  \\ 
m^T            & M'_N   \\
                \end{array} \right )
\end{equation}
where
\be
\label{U2}
\Omega= \begin{pmatrix}
 U             & 0    \\
 0             & W  \\
\end{pmatrix}, \;\;\; U^T Y_\nu {v\over \sqrt{2}} W = m, \;\;\; M_N' = W^T M_N W 
\ee
and
\begin{itemize}
\item $n_R=1$ 
\begin{equation}
m = \begin{pmatrix}
 0   \\
 0  \\
 m_D \\ 
\end{pmatrix}, \;\; m_D =  {v\over \sqrt{2}} \sqrt{\sum_{\alpha=1}^3 {Y_\nu}_\alpha^2}, \;\; U=U_{23}(\theta_{23}) U_{13}(\theta_{13}),\;\;\; W=1.
\end{equation}
\item $n_R=2$
\begin{equation}
m = \begin{pmatrix}
 0  & 0 \\
 0  & m_{D^-} \\
m_{D^+}&  0  \;\;\;
\end{pmatrix}, 
m_{D\pm}
={v \over 2} \left[\sum_{\alpha,i}{Y_\nu}_{\alpha i}^2\pm \sqrt{\left(\sum_{\alpha} {Y_\nu}_{\alpha1}^2-{Y_\nu}_{\alpha2}^2\right)^2+4\left(\sum_{\alpha} {Y_\nu}_{\alpha1}{Y_\nu}_{\alpha2}\right)^2}\,\right]^{1/2}
\end{equation}
\begin{equation}
U=U_{23}(\theta_{23}) ~U_{13}(\theta_{13})~ U_{12}(\theta_{12}), \;\;\; W(\theta_{45}).
\end{equation}
\end{itemize}
$U_{23}, U_{13}, U_{12}$ are $3\times 3$ orthogonal matrices corresponding to rotations around the 1, 2 and 3 axes and the 2$\times$2 $W$ matrix depends only on one extra angle, $\theta_{45}$.  This procedure can be easily extended to the case $n_R=3$, in which case 
$W$ would depend on three angles. 

Note that the number of real free parameters of ${\mathcal M}'_\nu$ in the new basis (the angles included in the matrices $U$ and $W$ and the mass parameters in $m$ and $M_N$) complete the expected total number and provide therefore a complete parametrization. For the CP phases, it can be shown that  one of them can be absorbed in a $\delta$-type phase in $U$, and  2/5 additional phases can be absorbed in $W$ for $n_R=2/3$ respectively. The latter disappear in the Dirac case, $M_N = 0$.  

We have not yet brought the full matrix to a diagonal form, but the necessary additional rotation angles must be calculable in terms of  the elements of  ${\mathcal M}'_\nu$.  We first note that the Dirac case, $M_N =0$, is trivial  for any $n_R$, because the final diagonalization is just a trivial rotation of angle $\pi/4$ (i.e. Majorana to Dirac basis) in the subsectors 34 for $n_R=1$,  in the sectors 34 and 25 for $n_R=2$ and 34, 25, 16 for $n_R \geq 3$.

Let us consider the diagonalization in the remaining cases.

\subsection{3+1 Model}

 For the simplest case $n_R=1$, the final diagonalization requires a 2$\times$2 rotation on the 34 subspace:
\begin{equation}
\label{massmatrix}
{\cal M_\nu}'' = \Omega'^T    {\cal M_\nu}' \Omega'  =   {\rm diag}(0,0,\lambda_-,\lambda_+),
\end{equation}
where
\begin{equation}
\lambda_-\equiv \frac{1}{2}\left(M-\sqrt{M^2+4m_D^2} \right) , \;\; \lambda_+\equiv \frac{1}{2}\left(M+\sqrt{M^2+4m_D^2} \right)
\end{equation}
and
\begin{equation}  
\;\;\;\; \Omega' =  \begin{pmatrix}
 1& 0 & 0 & 0\\
0 & 1 & 0 & 0 \\
0&  0 & \cos\theta_{34}  & \sin\theta_{34} \\ 
0&  0  & -\sin\theta_{34} & \cos\theta_{34} 
\end{pmatrix}, \;\;\; \sin  \theta_{34} = -\left[ {1\over 2} - {M\over 2 \sqrt{M^2 + 4 m_D^2}} \right]^{1/2}
\end{equation}
The full diagonalization matrix is therefore $\Omega~\Omega'$, where the  independent parameters are chosen to be $(m_D, M, \theta_{13}, \theta_{23})$. 

Obviously we could also use the completely general phenomenological parametrization with a $4\times 4$ mixing matrix, which relates the mass and flavour eigenstates, $U_{\rm mix}$. Such matrix is given, in terms of our parametrization by
\be
U_{\rm mix} = \Omega ~\Omega'.
\ee

\subsection{3+2 degenerate case}

If the Majorana mass matrix is proportional to the identity, $M_N = M I$, the matrix $M_N' =M_N$. The diagonalization of the full matrix involves  two independent $2 \times 2$ diagonalizations. We easily get
\bea
\label{eigenvalues3mas2}
{\cal M_\nu}'' = {\rm diag}(\lambda_0,\lambda_1,\lambda_2,\lambda_4,\lambda_3)
\eea
with 
\bea
\lambda_0 &=& 0 \nonumber\\
\lambda_1 &=& \frac{M}{2}-\sqrt{\left( \frac{M}{2}\right)^2+m_{D-}^2}\,,\nonumber\\ 
\lambda_2&=&\frac{M}{2}-\sqrt{\left( \frac{M}{2}\right)^2+m_{D+}^2}\,,\nonumber \\ 
\lambda_3&=&\frac{M}{2}+\sqrt{\left( \frac{M}{2}\right)^2+m_{D-}^2}\,,\nonumber \\ 
\lambda_4&=&\frac{M}{2}+\sqrt{\left( \frac{M}{2}\right)^2+m_{D+}^2}\,,\nonumber \\ 
\eea
where $\Omega'$ is a rotation on the sector 34 and 25 with angles
\bea
\label{t34t25}
\sin\theta_{34}&=&
-\left[ \dfrac{1}{2}-\dfrac{M}{2\sqrt{ M^2+4m_{D+}^2}}\right]^{1/2},\nonumber\\
\sin\theta_{25}&=&
\left[ \dfrac{1}{2}-\dfrac{M}{2\sqrt{ M^2+4m_{D-}^2}}\right]^{1/2}.
\eea
The free parameters in this case are $(M, m_{D^-},m_{D^+},\theta_{12},\theta_{13},\theta_{23})$. Notice that
here the rotation through the matrix $W$ is unphysical and consequently $\theta_{45}$ is not a 
physical parameter in the degenerate case.

It is easy to see how the Dirac and seesaw limits are obtained and how the parametrization reduces to the standard one in the three-neutrino mixing scenario. In the Dirac limit, the square neutrino mass matrix is $Y_\nu Y_\nu^T v^2/2 = U m m^T U^T= U {\rm Diag}(0, m^2_{D^-},m^2_{D^+}) U^T$ and therefore the matrix 
$U$ is just the PMNS matrix. In the seesaw limit, the same is true, since the light neutrino mass matrix is  $Y_\nu M^{-1}_N Y_\nu^T v^2/2 = M^{-1} U {\rm Diag}(0, m^2_{D^-},m^2_{D^+}) U^T$ for $M_N= M I$.

The completely general phenomenological parametrization with a $5\times 5$ mixing matrix, where the relevant matrix elements will be $U_{\alpha i}$, with $\alpha= e, \mu,\tau,..$ and $i=1,2,3,4,5$ would be, again,
\be
U_{\rm mix} = \Omega ~\Omega'.
\ee

\subsection{3+2 general case}

The final diagonalization of ${\mathcal M}'_\nu$ involves solving the eigensystem of a $2 n_R\times 2 n_R$ matrix, which is not possible analytically in the general case for $n_R > 1$. 

A simplification takes place in the seesaw limit ($Y_\nu v \ll M_N$), but in order for it to be simple it is necessary to first block diagonalize ${\mathcal M}_\nu$ to separate the light and heavy sectors. We can do this by taking the orthogonal rotation
\be
\label{U4}
\Omega=
\text{exp}\begin{pmatrix}
  0                    & \theta                     \\
-\theta^\dagger         & 0                     \\
\end{pmatrix} 
\,,
\ee
where the mixing $\theta$ is small in the seesaw limit. At leading-order 
\be
\theta = Y_\nu {v\over \sqrt{2}} M^{-1}_N,
\ee  
which satisfies
\bea
{ \mathcal M}'_\nu &=& \begin{pmatrix}
  I                    & -\theta                     \\
\theta^T         & I                    \\
\end{pmatrix} \begin{pmatrix}
  0                    & Y_\nu  v/\sqrt{2}               \\
Y_\nu^T  v/\sqrt{2}         & M_N                   \\
\end{pmatrix} \begin{pmatrix}
  I                    &\theta                     \\
-\theta^T         & I                    \\
\end{pmatrix}\\
&=& \begin{pmatrix}
  -Y_\nu M_N^{-1} Y_\nu^Tv^2/2   & 0                   \\
0   &        M_N  (1+{\mathcal O}(\theta^2)  )       \\
\end{pmatrix}. \nonumber\\
\eea
We can finally perform a rotation to diagonalize the light sector, which can be done by 
\be
{ \mathcal M}''_\nu = \Omega'^T { \mathcal M}'_\nu \Omega'\;\;\; \Omega' = \begin{pmatrix} \tilde{U} & 0 \\
0 & I 
\end{pmatrix},
\ee
where $\tilde{U}$ can be chosen such that
\be
\tilde{U}^T Y_\nu M_N^{-1/2} \tilde{W} v/\sqrt{2} = m^{1/2}\,.
\label{eq:ci}
\ee
$\tilde{U}$ can be identified with the usual PMNS matrix (up to phases) while $\tilde{W}$ is a 
2$\times 2$ orthogonal matrix that depends on an extra mixing angle that we can call $\theta_{45}$. $m^{1/2}$ contains the two non-zero light neutrino masses:
\begin{equation}
\label{eq:msqrt}
m^{1/2} = \begin{pmatrix}
 0  & 0 \\
 0  & m_2^{1/2} \\
m_3^{1/2}&  0  \;\;\;
\end{pmatrix}. 
\end{equation}
 We can choose as mass parameters the two masses in 
$m^{1/2}$ and the $n_R$ heavy masses in $M_N$. In addition, we have the 3 standard angles in 
$\tilde{U}(\theta_{23},\theta_{13},\theta_{12})$ and the extra angle of 
$\tilde{W}(\theta_{45})$. 
It is important to stress that for the CP conserving case eq.~(\ref{eq:ci}) is the usual generic bidiagonalization of the matrix $Y_\nu M_N^{-1/2}$ while in the general case, it should be understood as a definition of the orthogonal matrix $\tilde{W}$. 
This parametrization turns out to be equivalent to  that of Casas-Ibarra \cite{Casas:2001sr}, where the matrix  $\tilde{W}$ is what they call the matrix $R$.
 
Note that the matrix $\Omega$ can now be reconstructed from 
these parameters since
\be
\theta = \tilde{U} m^{1/2} \tilde{W}^T M^{-1/2}_N.
\label{eq:magic}
\ee

The phenomenological parametrization in this case is 
\be
\label{eq:uphenodeg}
U_{\rm mix} = \Omega~ \Omega' \simeq \begin{pmatrix}
 \tilde{U}  & \theta \\
 -\theta^T  \tilde{U}  & I \;\;\;
\end{pmatrix} +{\mathcal O}(\theta^2),  
\ee
and therefore $(U_{\rm mix})_{\alpha i} = \theta_{\alpha i}$ for $\alpha=e,\mu,\tau$ and $i=4,5$. 

Obviously this parametrization is not the same as the one discussed before. However, 
note that in the degenerate case the matrices $U$ and $\tilde{U}$ coincide. 
This parametrization has the advantage to be more physical in the seesaw limit, but cannot be 
extrapolated to the quasi-Dirac limit. For a full exploration of the parameter space that works 
in the quasi-Dirac and seesaw limits, the general parametrization of eqs.~(\ref{U2}) is more appropriate. 

\section{Constraints from oscillations on 3+1 model}
\label{sec:3+1}

We start by considering the $3+1$ model. Although the model is very constrained, it has in principle sufficient parameters to fit 
two mass splittings and two mixing angles. Therefore, it is interesting to understand to what extent  the model can fit oscillation data. 

We have explored the full  parameter space trying to fit the data from long-baseline accelerator (MINOS \cite{Adamson:2011ig}) and 
reactor experiments (KAMLAND \cite{Gando:2010aa}, and CHOOZ\cite{Apollonio:1999ae}), since they provide the most precise information on the  oscillation frequencies.  We carry out the analysis of these experiments as explained in \cite{GonzalezGarcia:2010er,GonzalezGarcia:2011my}, and we refer to those papers for details. 

\begin{figure}[!ht]
\begin{center}
\includegraphics[width=14cm]{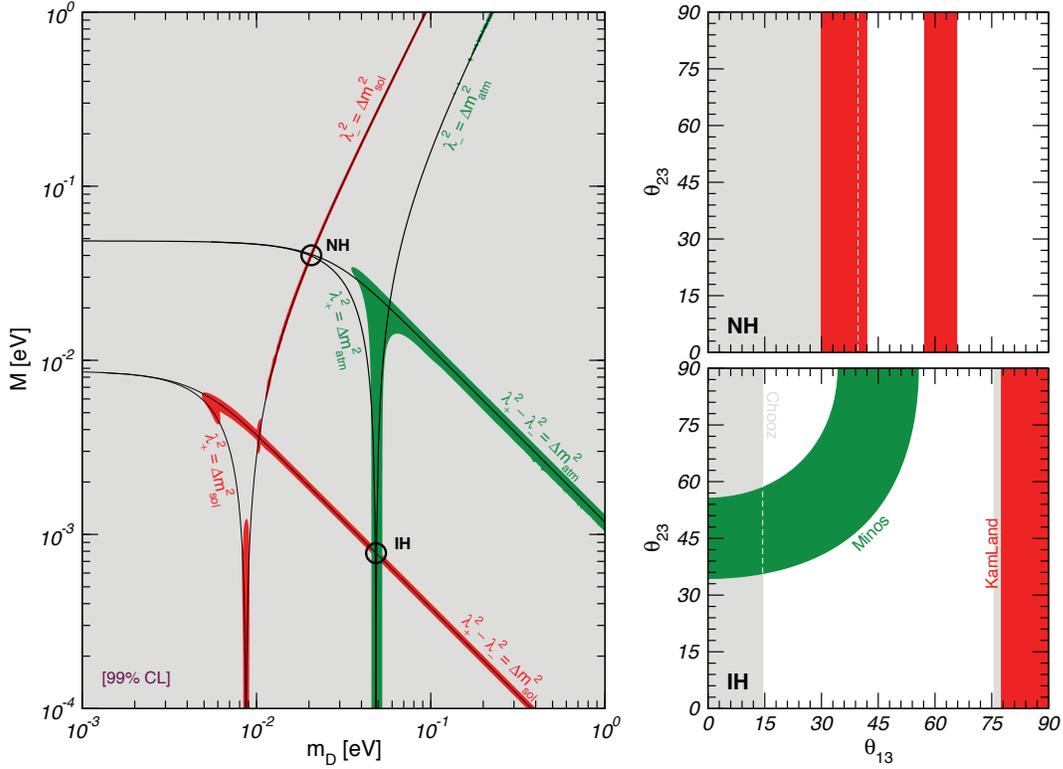}
\vspace{0.5cm}
\end{center}
\caption{Left: Contours corresponding to $\chi^2_{exp}= \chi^2|^{SM}_{\rm exp} + 9.21$ from the fits of the 3+1 model to KAMLAND (Red), MINOS (Green) and CHOOZ (grey) on the plane $(m_D, M)$ (after minimizing in the angles). The solid lines correspond to the values of $m_D$ and $M$ such that one of the mass square differences in the spectrum coincides with the solar or atmospheric splitting. The circles indicate the values of $m_D, M$ where the solar and atmospheric splittings coexist for the normal (NH) and inverted hierarchy (IH) . Right: same contours on the plane $(\theta_{13}, \theta_{23})$, for the values of $m_D, M$ fixed at the intersection points NH and IH. }
\label{fig:3+1}
\end{figure}

On the left plot of Fig.~\ref{fig:3+1}, we show the contours $\chi^2_{\rm exp}= \chi^2_{SM} + 9.21$ on the plane of the two mass parameters $(m_D, M)$ after minimizing in the two angles. $\chi_{SM}^2$ corresponds to the best fit of the standard three-neutrino oscillation scenario, while $\chi^2_{\rm exp}$ corresponds to  the minimum  in the fits of the $n_R=1$ model to the experiments (exp): KAMLAND (red), MINOS (green) and CHOOZ (grey). As expected MINOS and KAMLAND allow a very constrained region of parameter space that deviates very little from the lines where one of the mass splittings equals $\Delta m^2_{atm}$ and $\Delta m^2_{sol}$ respectively. The two lines intersect only in the two points indicated by circles, for the normal (NH) and inverted (IH) hierarchy. On those points, the atmospheric and solar frequencies can be accommodated simultaneously, but the values of the angles that are required to fit the data are nevertheless incompatible, as shown on the two right panels. 

The conclusion is therefore that the $3+1$ is excluded. Note however that the combination of the three experiments is essential. This is in agreement with the qualitative analysis of  ref.~\cite{Liao:2005hr}.

\section{Constraints from oscillations on 3+2 model}
\label{sec:3+2}
For the case $n_R=2$, the number of parameters is quite large to do a full-fledged fit as has been done in the $3+1$ model. We will consider the simpler degenerate limit, where there are only three independent masses and three angles and furthermore consider the CP conserving case. In order to identify the allowed regions we will follow the strategy of looking for solutions where at least one of the mass splittings corresponds to the atmospheric one and another one to the solar. Given that both splittings are well established by data, this method should be rather robust in pinning down local minima. 

\begin{figure}[!ht]
\begin{center}
\includegraphics[width=10cm]{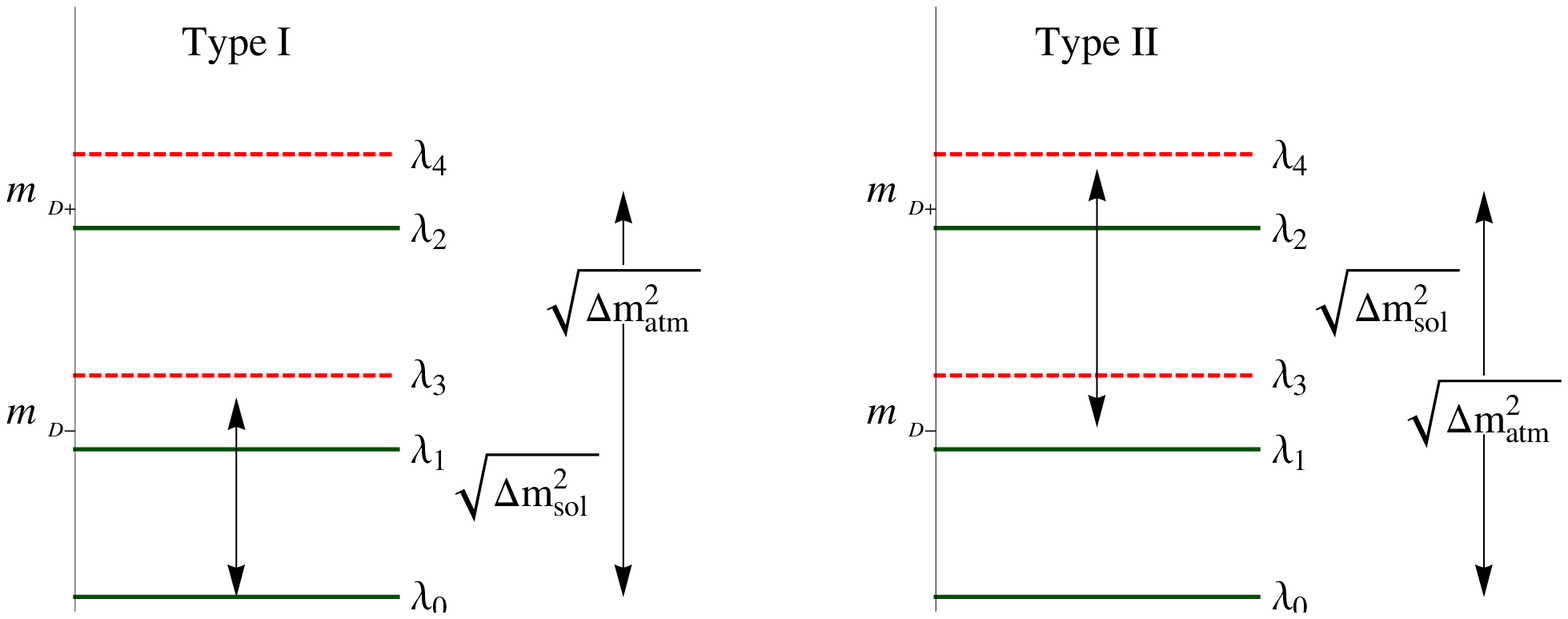}
\includegraphics[width=10cm]{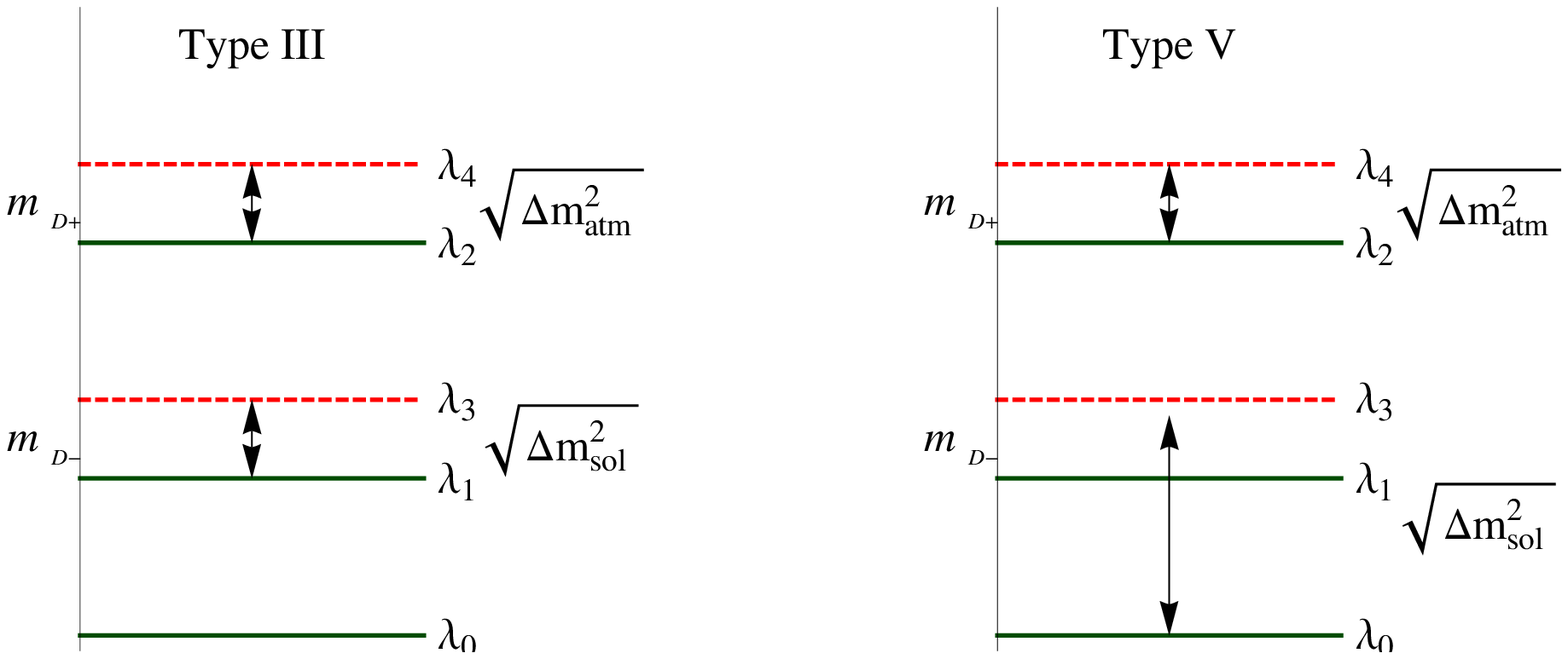}
\includegraphics[width=10cm]{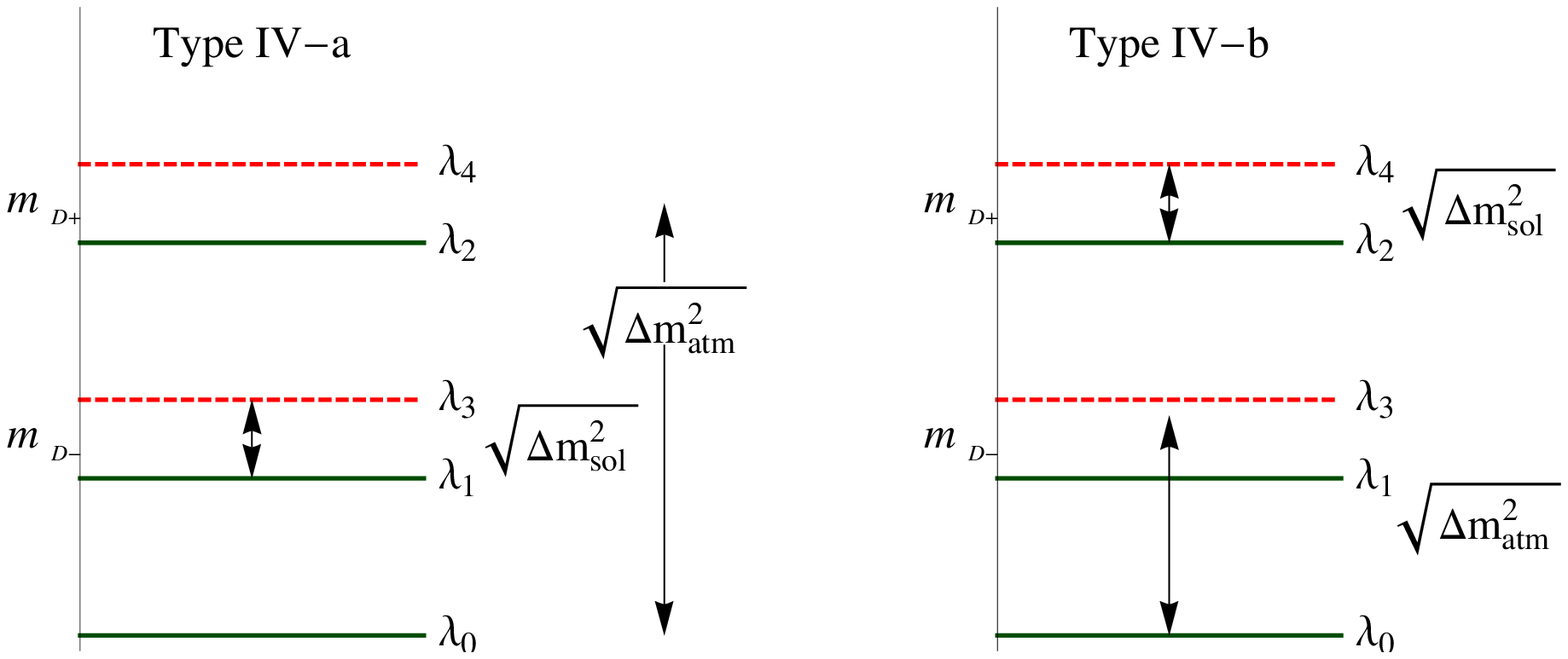}
\vspace{0.5cm}
\end{center}
\caption{Spectral structures that contain the solar and atmospheric splitting in the quasi-Dirac region. }
\label{fig:spectra}
\end{figure}

More concretely we consider all possible combinations $ijkl$ such that
\begin{eqnarray}
\left|\lambda^2_{i}-\lambda_j^2\right| = \Delta m^2_{\rm atm} = 2.5 \times 10^{-3} {\rm eV}^2, \;\;\; \left|\lambda^2_{k}-\lambda^2_l\right| = \Delta m^2_{\rm sol} = 8 \times 10^{-5} {\rm eV}^2,  \;\;\; i,j,k,l=0,..,4\nonumber\\
\end{eqnarray}
These two equations imply relations between $M$,  $m_{D^+}^2$ and $m_{D^-}^2$ that we solve numerically for $m_D^+$ and $m_D^-$ as a function of $M$. 
In principle there could be up to 90 combinations, but we restrict to those that correspond to $m_{D^+} ^2 > m_{D^-}^2$, since they have  been ordered in this way. In the quasi-Dirac limit $M \ll m_{D^+}, m_{D^-}$, most combinations are possible, while in the opposite limit 
$M \gg m_{D^\pm}$ two states decouple and therefore only the standard choices of the $3\nu$ scenario survive (two for the normal hierarchy and two for the inverse one). We can therefore classify all the possible solutions  by the hierarchy they represent in the quasi-Dirac limit. We find the following, qualitatively distinct,   spectral patterns depicted in Fig.~{\ref{fig:spectra}}. 

\begin{itemize}

\item Type I: Quasi-Dirac Normal Hierarchy

For small $M$, we could have 
the solar splitting  between eigenstates 0 and 1 or 0 and 3, and it is therefore related to $m_{D^-}\sim \sqrt{\Delta m^2_{\rm sol}}$. The atmospheric splitting could be either between 0 and the heavier states 2 or 4, ie. $m_{D^+}\sim \sqrt{\Delta m^2_{\rm atm}}$. But it could also be  between the states 1/3 and 2/4. There are in total 12 combinations that differ only by small perturbations of  ${\mathcal O}(\Delta m^2_{\rm sol}/\Delta m^2_{\rm atm})$. Typically these solutions have other mass splittings that are much smaller than both solar and atmospheric for sufficiently small $M$. As  $M$ increases, most of these solutions disappear except the two that involve only the states 0,1,2.


\item Type II: Quasi-Dirac Inverse Hierarchy 

This case is like type I but representing an inverted hierarchy. This happens when the atmospheric mass splitting occurs between states 0 and 1/3 or 2/4, while the solar one does between states 1/3 and 2/4. In this case there are 16 combinations in total, which have 
$m_{D^-} \simeq \sqrt{\Delta m^2_{\rm atm}}$ and $m^2_{D^+} - m^2_{D-} \simeq  \sqrt{\Delta m^2_{\rm sol}}$. Again most of these solutions do not survive for large $M$, only two do. All the additional splittings are not larger than those measured. 


\item Type III:  Quasi-Dirac Degenerate 

In this case, the solar splitting is the one between states 1 and 3, while the atmospheric one is the splitting between states 2 and 4. This solution is only possible for not too large $M$ and it is rather peculiar because it has other splittings that are larger than those measured,  since typically $m_{D^+}, m_{D^-} \gg \sqrt{\Delta m^2_{\rm atm}}, \sqrt{\Delta m^2_{\rm sol}}$. It is extremely interesting if it gives a good fit to the data. 


\item Type IV:  Quasi-Dirac Solar Degenerate 

This is a mixed situation between the quasi-Dirac  and quasi-degenerate cases. There are two possibilities. In type IV-a,  the solar splitting is the one between states 1 and 3, while the atmospheric one is driven by $m_{D^+}$, that is the splitting between states 0 and 2 or 4 or alternatively 
1 or 3 and 2 or 4. There are 6 possibilities of this type. Type IV-b corresponds to a solar splitting associated to the states 2 and 4. Then the atmospheric must be related to $m_{D^-}$, so it is the difference between states 0 and 1 or 3. Two combinations exist of this type.



\item Type V:  Quasi-Dirac Atmospheric Degenerate 

 In this case, the atmospheric splitting is the one between states 
2 and 4, while the solar one is between states 0 and 1 or 3. There are two possible combinations of this type, which correspond to 
a situation with $m_{D^-} \sim \sqrt{\Delta m^2_{\rm sol}}$ and $m_{D^+} \gg m_{D^-}$.

\end{itemize}
There are some more solutions that only occur in the intermediate region of $M$ where no clear hierarchies can be established. 
These however give a very poor fit to the data, so we will not discuss them further.

\subsection{Fits to reactor and accelerator experiments}

We started by considering the fits to long-baseline reactor and accelerator experiments (LBL), that is including Chooz, KamLAND, MINOS appearance and disappearance for neutrinos and antineutrinos, that in principle have the most precise information about the two mass splittings.  We find that solutions of all the types mentioned provide a good fit to the data, as good as the standard three neutrino scenario. That this holds is obvious in two limits: the Dirac limit $M \rightarrow 0$, where the spectrum matches a three-neutrino mixing scenario with one massless neutrino, and, for those solutions that survive,  the seesaw limit $M \gg m_{D^\pm}$, where the spectrum also corresponds to a three-neutrino mixing scenario with one massless neutrino, since the two heavier states decouple. 
The survival of other solutions with $M$ between the two limits is however non-trivial since they imply the presence of other 
mass splittings that could affect oscillations.  

\begin{figure}[!ht]
\begin{center}
\includegraphics[width=7cm]{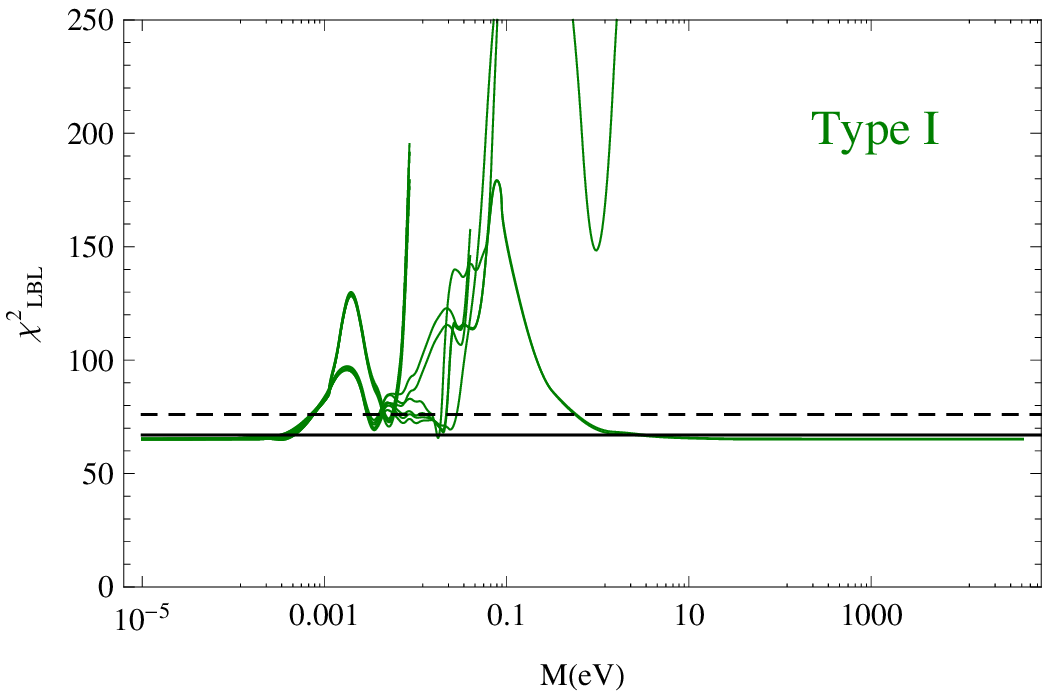} \includegraphics[width=7cm]{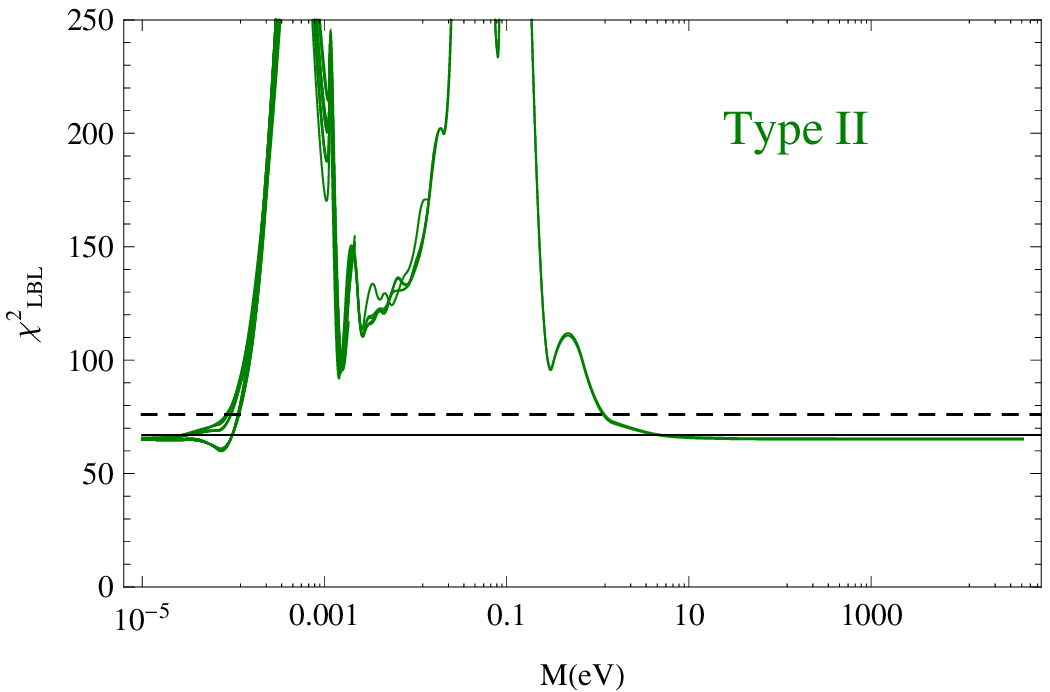}
\vspace{0.5cm}
\includegraphics[width=7cm]{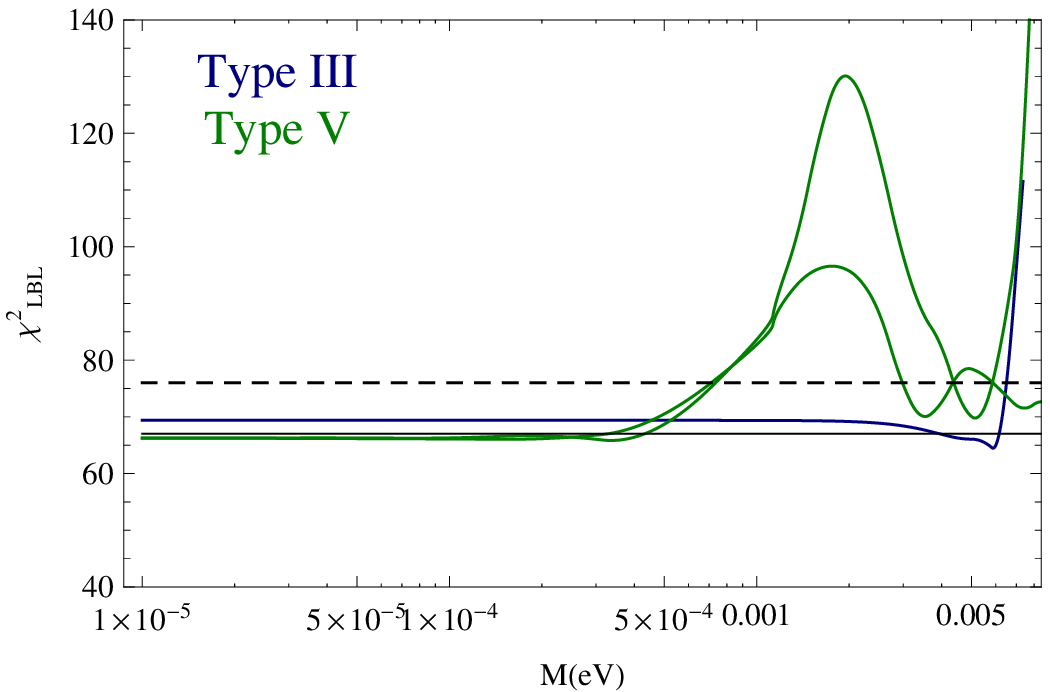} \includegraphics[width=7cm]{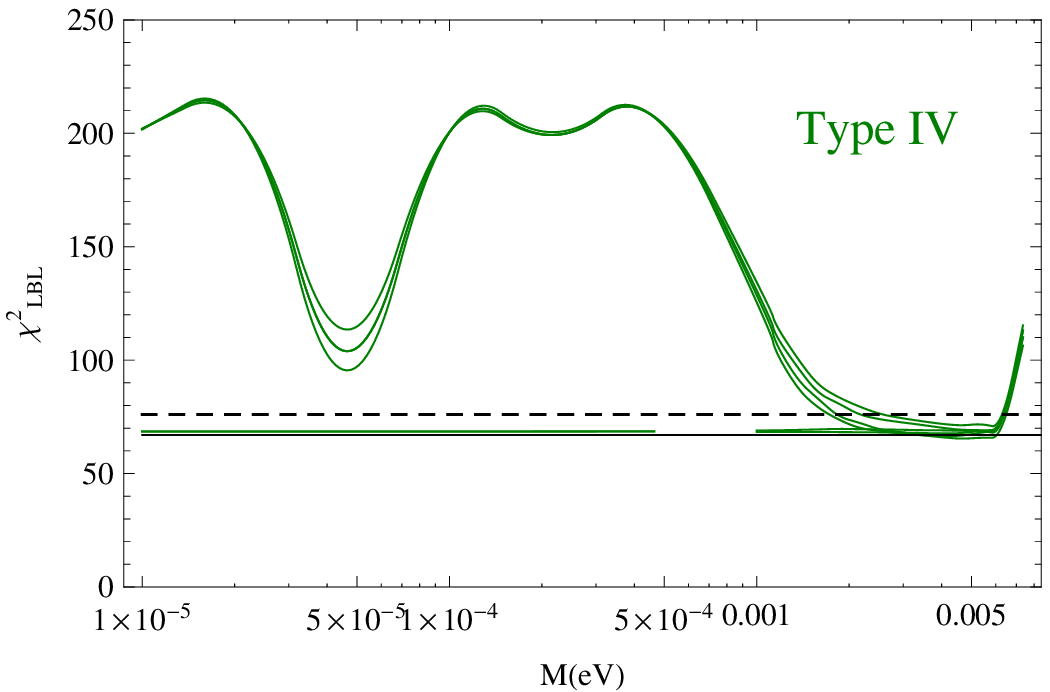}
\end{center}
\caption{Minimum $\chi^2_{LBL}$ from the fits to KamLAND, MINOS and Chooz as a function of $M({\rm eV})$. The solid(dashed) horizontal line corresponds to $\chi^2_{SM}$($\chi^2_{SM}+9$) of the standard 3$\nu$ scenario. Up left: solutions of type I. Up right: solutions of type II. Down left: solutions of types III and V. Down right: solutions of type IV.}
\label{fig:lbl}
\end{figure}

In Figs.~\ref{fig:lbl} we show the minimum $\chi^2$ after minimization in all the angles, as a function of $M$. We only show those solutions that give a good fit for some value of $M$, which can be classified according to the hierarchies above.

The left upper plot corresponds to the solutions of type I . As expected the $\chi^2$ increases significantly with respect to the Dirac-limit value for intermediate values of $M$. From the point of view of LBL data we see that values in the range $5 \times10^{-4}{\rm eV} \lesssim M\lesssim 1$eV give poor fits to the data. There are however also some values of $M$ in this range that give acceptable fits to the data for some of the solutions, that need further scrutiny.

A slightly different situation is found for the solutions of type II, which for small/large enough $M$ correspond to an inverse hierarchy.  Here LBL data can exclude a slightly  wider region $10^{-4}$ eV $\lesssim M \lesssim 1$eV and we do not see any local minimum that provides a good fit within this range.

The quasi-degenerate and mixed solutions are shown in the lower plots. In this case the solutions only exist for sufficiently small values of $M \lesssim 5 \times 10^{-3}$eV, as shown. The plot on the left corresponds to types III and V, while the one on the right corresponds to types IV. The type III gives a rather good fit to the LBL data in all the region, while solutions of type V only for $M \lesssim 5 \times 10^{-4}{\rm eV}$ or for $M \simeq 0.005$~eV. Some of the solutions of type IV instead give a good fit in most of the range where they exist: $10^{-5} {\rm eV} \lesssim M \lesssim 10^{-3}$eV.

\begin{figure}[!ht]
\begin{center}
\includegraphics[width=12cm]{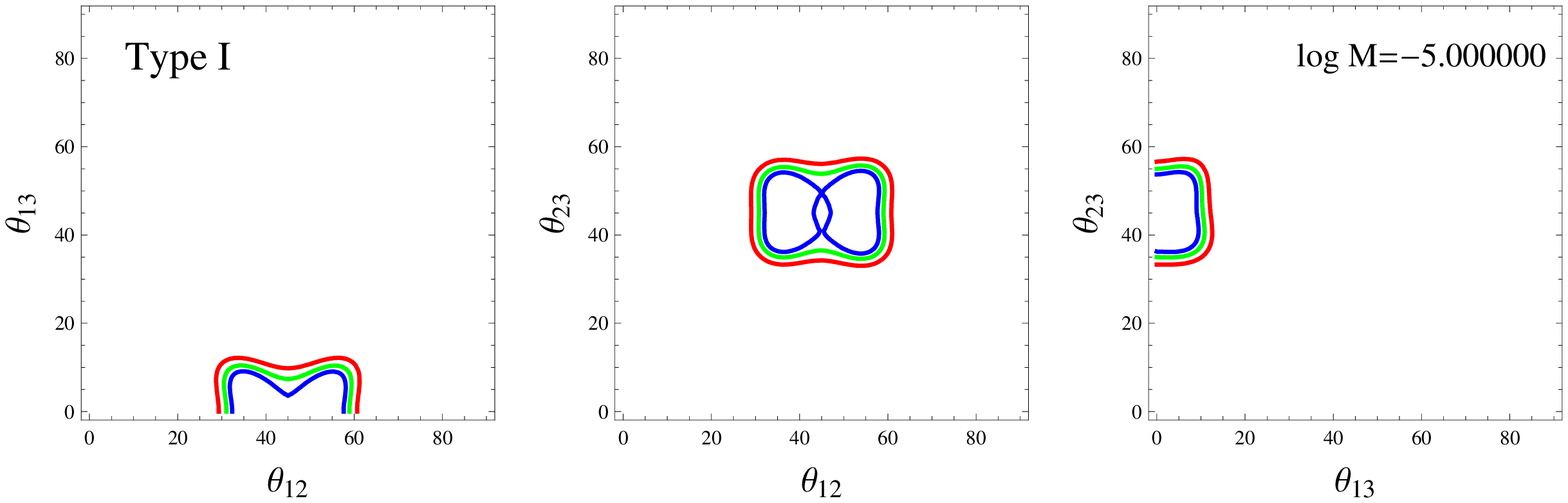} 
\includegraphics[width=12cm]{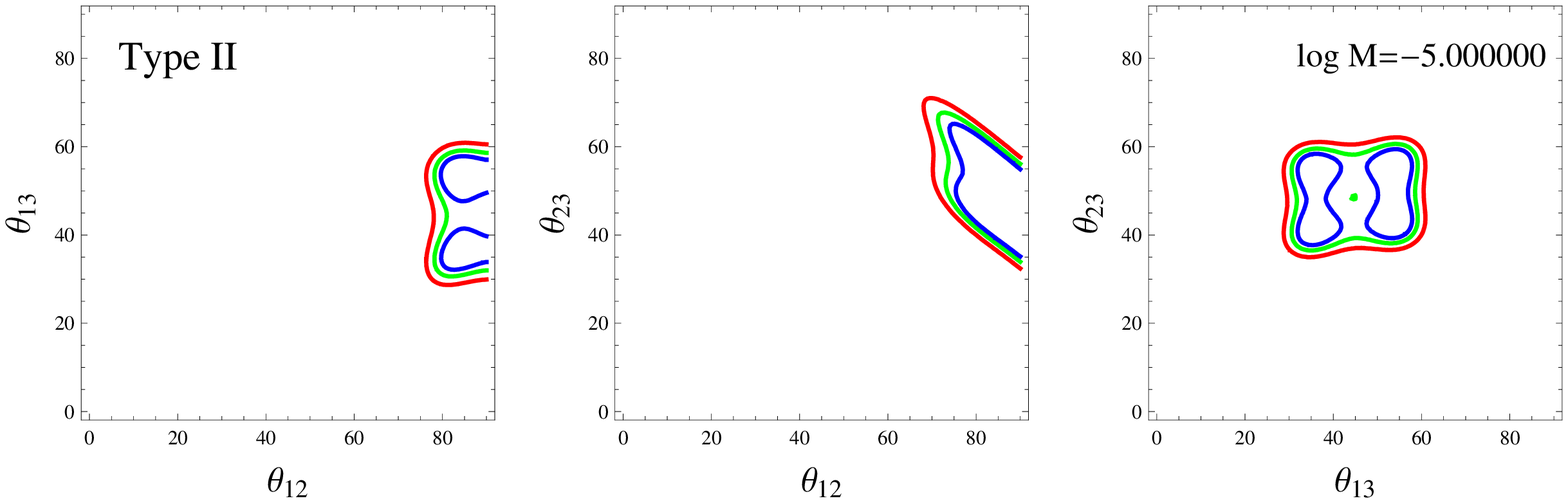}
\end{center}
\caption{$\chi^2_{LBL}$ contours (1,2,3$\sigma$) on the planes $(\theta_{12},\theta_{13})$, $(\theta_{12},\theta_{23})$ and $(\theta_{13},\theta_{23})$ at fixed $M$ as shown, and  for solutions of type I (up) and II (down).}
\label{fig:angles}
\end{figure}

In Figs.~\ref{fig:angles} we show the minimum $\chi^2_{LBL}$ on the planes defined by each pair of mixing angles for the solutions of type I and II as examples. Typically one minimum is found, but there are octant degeneracies, as expected. The solutions can be easily identified by the very different positions of their minima on these planes. 

 Note that we  consider the same ordering of states in all cases and therefore the solar and atmospheric mass splittings are labelled differently for the different cases. As a result, only in the normal hierarchy cases the values of the angles $\theta_{12}$, $\theta_{13}$ and $\theta_{23}$ are in one-to-one correspondance with those in the standard parametrization of the three neutrino scenario. In the inverse hierarchy case, the angles are quite different, although the oscillation probabilities in vacuum will be  equivalent. For example,  the normal hierarchy for the angles $(\theta_{12}=\theta_{\rm sol},\theta_{13}=0,\theta_{23}=\theta_{\rm atm})$ is equivalent for the inverse hierarchy to $(\theta_{12}=\pi/2,\theta_{13}=\theta_{\rm sol},\theta_{23}=\pi-\theta_{\rm atm})$, as can be seen in Figs.~\ref{fig:angles}.

In summary, LBL data are able to exclude most solutions in an intermediate region of $M$, but some peculiar solutions do survive. As expected, the quasi-Dirac type I and II solutions survive for all $M \leq M^{QD}_{max}$, and those involving only the lighter 012 states
also survive in the seesaw limit, $M\geq M^{SS}_{min}$. 

 In the following we will further constrain  the intermediate solutions using solar and atmospheric data, and also refine the determination of the quasi-Dirac and seesaw boundaries: $M_{max}^{QD}$ and $M_{min}^{SS}$.

\subsection{Solutions for intermediate $M$}

\begin{figure}[!ht]
\begin{center}
\includegraphics[width=16cm]{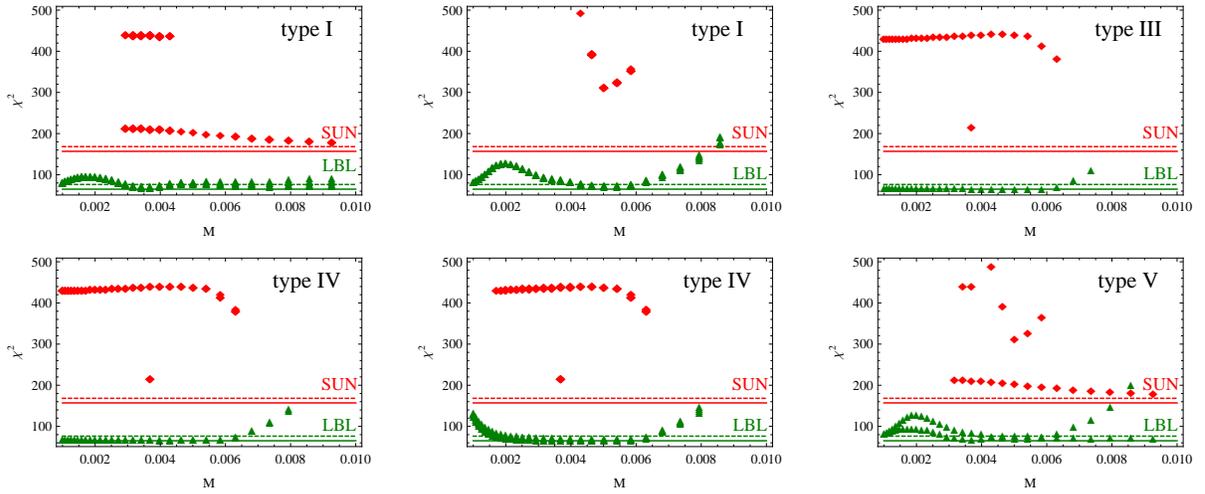}
\end{center}
\caption{ $\chi^2_{LBL}$ (green triangles) from the fits to KamLAND, MINOS and Chooz and $\chi^2_{SUN}$ (red diamonds) as a function of $M({\rm eV})$. The  horizontal lines corresponds to the $\chi^2_{min}$ of the standard 3$\nu$ scenario for the LBL (green) and solar (red) fits. The dashed lines correspond to a shift by $\Delta\chi^2=9$.}
\label{fig:mev}
\end{figure}

\begin{figure}[!ht]
\begin{center}
\includegraphics[width=15cm]{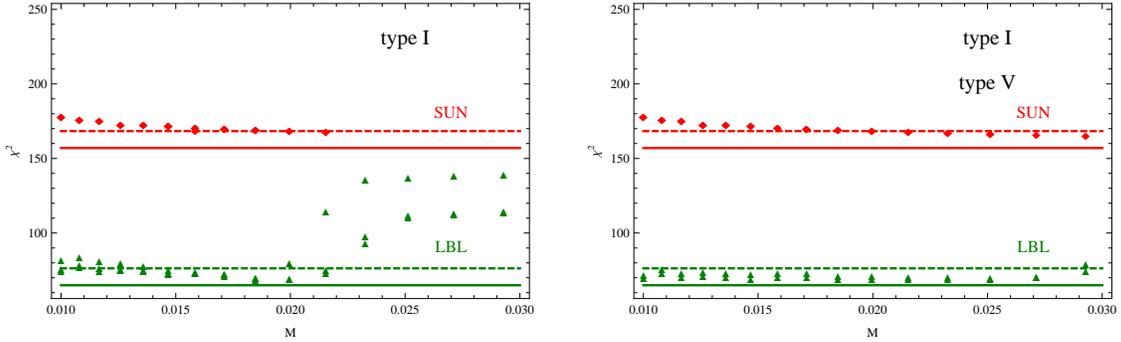}
\end{center}
\caption{$\chi^2_{LBL}$ (green triangles) from the fits to KamLAND, MINOS and Chooz and $\chi^2_{SUN}$ (red diamonds) as a function of $M({\rm eV})$. The  horizontal lines corresponds to the $\chi^2_{min}$ of the standard 3$\nu$ scenario for the LBL (green) and solar (red) fits. The dashed lines correspond to a shift by $\Delta\chi^2=9$.}
\label{fig:cev}
\end{figure}

Solar data give additional constraints on solutions in the intermediate region (we refer to \cite{GonzalezGarcia:2010er} for details of the data analysis). The crucial piece of information provided by solar data is of course the neutral current measurement of SNO \cite{Aharmim:2009gd}.  In Fig.~\ref{fig:mev} we compare the $\chi^2 $ of the LBL fit and that of solar data to those of the standard $3\nu$ scenario (horizontal lines) for $M$ in the meV range, $10^{-3}$--$10^{-2}$eV. All solutions that survive LBL are essentially excluded by solar data.
The same is shown in Fig.~\ref{fig:cev} for the ceV region, $10^{-2}$--$10^{-1}$eV. 
In this case, a small region around $M\gtrsim 10^{-2}$eV survives both fits for types I and V. For those cases, we have considered the fit to atmospheric data \cite{Wendell:2010md} and the result is that none provide a decent fit to atmospheric data and are therefore excluded.

\subsection{Limit on quasi-Dirac neutrinos}

\begin{figure}[!ht]
\begin{center}
\includegraphics[width=10cm]{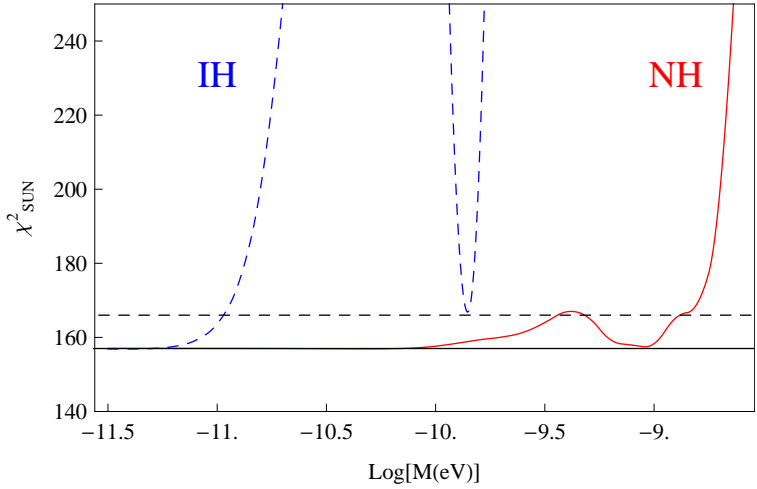}
\vspace{0.5cm}
\end{center}
\caption{Minimum $\chi^2_{SUN}$ from solar data as a function of $M({\rm eV})$ in the quasi-Dirac region for solutions of type I (NH) and of type II (IH). The solid horizontal line corresponds to the minimum $\chi^2$ and the dashed one to a shift by $\Delta \chi^2= 9$. }
\label{fig:sunqd}
\end{figure}

Solutions of types I and II provide a perfect fit to the data for $M \lesssim M_{max}^{QD}$, which bounds the quasi-Dirac region. Including solar data is essential in constraining $M_{max}^{QD}$ to much lower values that those obtained from just fitting LBL data,  for which $M_{max}^{QD} \sim 5 \times 10^{-4}$ eV (see Fig.~\ref{fig:lbl}). The upper bound $M_{max}^{QD}$ is therefore constrained exclusively by solar data. The result is shown in Fig.~\ref{fig:sunqd} for the solutions that have a Dirac limit, i.e. those of types I and II. We obtain a value of $M_{max}^{QD} \simeq 10^{-9},10^{-11}$ eV for the NH, IH solutions respectively. In the last section we estimate analytically such extremely small limits. 

\subsection{Mini-seesaw region}

\begin{figure}[!ht]
\begin{center}
\includegraphics[width=10cm]{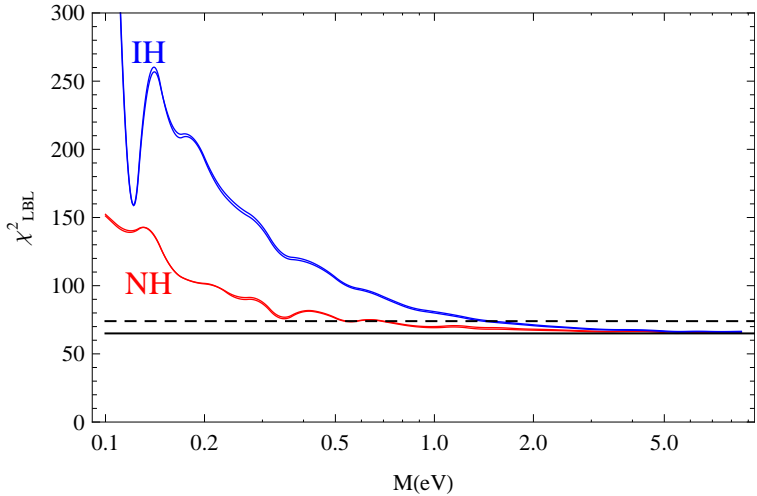}
\vspace{0.5cm}
\end{center}
\caption{Minimum $\chi^2_{LBL}$  as a function of $M({\rm eV})$ in the mini-seesaw region  for solutions of type I (NH) and of type II (IH). The solid horizontal line corresponds to the minimum $\chi^2$ and the dashed one to a shift by $\Delta \chi^2= 9$. }
\label{fig:mss}
\end{figure}

For $M$ sufficiently large, the states 3 and 4 decouple from the spectrum and only four solutions provide good fits to LBL data (two corresponding to NH and two to IH). The value at which this seesaw limit sets in is $M_{min}^{SS}$ and it is essentially determined by LBL data, as shown in Fig.~\ref{fig:mss}. We find $M_{min}^{SS} \simeq 0.6$ eV, 1.4 eV for the NH, IH solutions, respectively. 

\begin{figure}[!ht]
\begin{center}
\includegraphics[width=14cm]{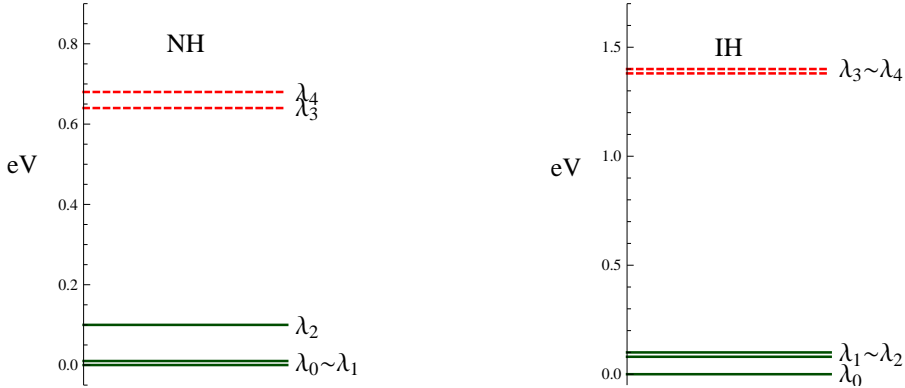}
\vspace{0.5cm}
\end{center}
\caption{Spectra of neutrino mass eigenstates in the mini-seesaw region for  $M \sim M^{SS}_{\rm min}$, $M\sim 0.6$eV for the normal (left) and $M \sim 1.4$~eV for the inverse hierarchy (right).}
\label{fig:mss_hier}
\end{figure}

Neither solar nor atmospheric data  add any further constraint on $M_{min}^{SS}$. The mini-seesaw solutions provide a perfect fit to solar and atmospheric data for larger values of $M$. Near the seesaw threshold the mini-seesaw solutions that agree with solar, atmospheric and LBL data have a hierarchy of the form shown in Figs.~\ref{fig:mss_hier}.
 
\section{Other bounds in the mini-seesaw region}
\label{sec:lsnd}

\subsection{LSND/MiniBoone and reactor}

Obviously, an interesting question remains concerning the neutrino anomalies: the old LSND result \cite{Aguilar:2001ty}, recently reinforced by the anti-neutrino appearance data in MiniBoone \cite{mb2,mb1}, the low-energy excess of MiniBoone and the recently discovered reactor neutrino anomaly \cite{Mueller:2011nm,Mention:2011rk} . A possible explanation of such anomalies is the existence of extra sterile neutrinos  in the eV range \cite{Akhmedov:2010vy,Nelson:2010hz,Kopp:2011qd}. It is then an obvious question whether our 3+2 model for $M > M_{min}^{SS}$ can account for any of those anomalies. 

In \cite{Kopp:2011qd} a recent reanalysis of the 3+2 phenomenological scheme shows that there is a significant improvement in the goodness-of-fit with respect to the analysis previous to the reevaluation of the  reactor fluxes, and to the publication of the MiniBoone antineutrino data. Nevertheless, there is still a  
tension between the LSND/MiniBoone appearance and the muon disappearence experiments. 
  
According to \cite{Kopp:2011qd}, a good fit to the data is obtained with:
\begin{enumerate}
\item  
Two neutrino mass differences around $1$ eV (more precisely around $0.5$ eV and $0.9$ eV).
\item  
In the phenomenological parametrization, large enough $|U_{e4}|$, $|U_{e5}|$, $|U_{\mu 4}|$ and $|U_{\mu 5}|$, around $10^{-1}$. 
\item   
CP-violation.  
\end{enumerate}
Since the parametrization for the phenomenological 3+2 model contains all the possible physical parameters when there are 
5 mass eigenstates, our model can be described by the same parametrization although there are correlations between the parameters, in particular between the masses and angles. In particular for the degenerate case, according to eq.~(\ref{eq:uphenodeg}), we have
\bea
\label{matrixelements}
(U_{\rm mix})_{e4}&=& s_{13}s_{34},\nonumber\\
(U_{\rm mix})_{e5}&=& c_{13}s_{12}s_{25},\nonumber\\
(U_{\rm mix})_{\mu 4}&=& c_{13}s_{23}s_{34},\nonumber\\
(U_{\rm mix})_{\mu 5}&=& (c_{12}c_{23}- s_{12}s_{13}s_{23})s_{25},
\eea
with 
\bea
s_{25}
\approx m_{D^-}/M,\, \;\;\;\;
s_{34}
\approx -m_{D^+}/M,
\eea
for $M$ in the eV range. 

Concerning  the massive states, their masses are: $(\sim m^2_{D^-}/M, \sim m^2_{D^+}/M, \sim M, \sim M)$.
Therefore, after taking into account the results from the above sections, the only parameter we can play with is the
 Majorana mass $M$, and therefore in the degenerate case it is not easy to accommodate two distinct eV masses. On the other hand the matrix elements $U_{e4/5}$ are intriguingly in the right ballpark.

In order to have a more explicit prediction we have evaluated the $P_{ee}$ and $P_{e\mu}$ oscillation probabilities in the mini-seesaw regime analytically via a perturbative expansion in the small parameters
\begin{equation}
\epsilon_- = \left (\frac{m_{D_-}}{M} \right)^2, \,  \qquad
\epsilon_+ =\left (\frac{m_{D_+}}{M}\right)^2.
\end{equation}
At second order in $\epsilon_\pm$ we get for the ten mass differences: 
\begin{eqnarray}
\Delta m^2_{ij}= {\mathcal O}(\epsilon^2 M^2)  , ~\; \Delta m^2_{4i} \sim  \Delta m^2_{5i} \sim M^2 (1+{\mathcal O}(\epsilon^2)), ~ \Delta m^2_{54} \sim {\mathcal O}(\epsilon M^2) , ~ ~i,j=1,3. 
\end{eqnarray}

\begin{figure}[!ht]
\begin{center}
\includegraphics[width=10cm]{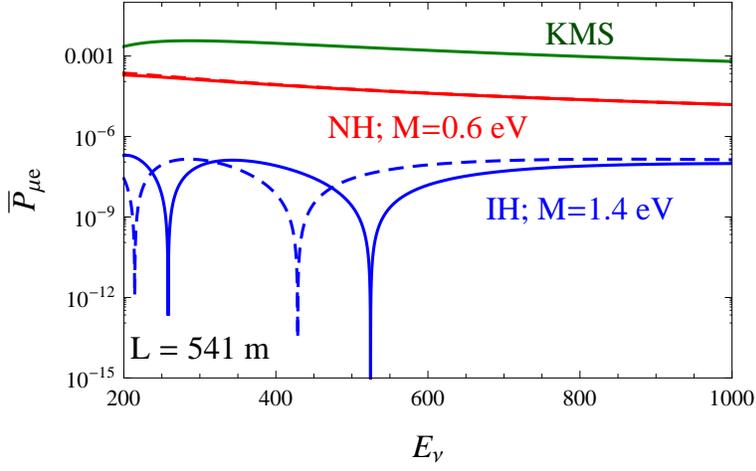}
\vspace{0.5cm}
\end{center}
\caption{ The $\bar{P}_{\mu e}$ antineutrino oscillation probability at $L = 541$ m as a function of $E_\nu$ for the NH (red, $M = 0.6$ eV) and IH (blue, $M = 1.4$ eV). The best fit result of \cite{Kopp:2011qd} is labelled KMS. The dashed lines correspond to the perturbative results.}
\label{fig:MB}
\end{figure}

At the same order in $\epsilon_\pm$ the mixing angles are: 
\begin{eqnarray}
\sin^2 \theta_{25} = \epsilon_- -3 \epsilon_-^2, \;\sin^2 \theta_{34} = \epsilon_+ -3 \epsilon_+^2, \nonumber
\end{eqnarray}
and finally the oscillation probability is
\begin{equation}
P_{e\mu}^{SS} = 4 \sin^2 \left ( \frac{M^2 L}{4 E_\nu}\right ) \left ( \epsilon^2_+ A_{44}  + 2 \epsilon_+ \epsilon_- A_{54} +\epsilon_-^2 A_{55}\right ) + O(\epsilon_\pm^3),
\end{equation}
where
\begin{eqnarray}
A_{44} &=& \cos^2 \theta_{13} \sin^2 \theta_{13} \sin^2 \theta_{23}, \nonumber \\
A_{54} &=& \cos^2 \theta_{13} \sin \theta_{12} \sin \theta_{13} \sin \theta_{23} (\cos \theta_{12} \cos \theta_{23} - \sin \theta_{12} \sin \theta_{13} \sin \theta_{23}), \\
A_{55} &=& \cos^2 \theta_{13} \sin^2 \theta_{12} (\cos \theta_{12} \cos \theta_{23} - \sin \theta_{12} \sin \theta_{13} \sin \theta_{23})^2.\nonumber
\end{eqnarray}
In the normal hierarchy case, $\theta_{13}$ is small and will introduce a suppression of the $A_{44}$ and $A_{54}$ coefficients with respect to $A_{55}$. For vanishing $\theta_{13}$ we get:
\begin{equation}
\label{eq:pemumSSDH}
P_{e\mu}^{SS,NH} = 4 \sin^2 \left ( \frac{M^2 L}{ 4 E_\nu}\right ) \epsilon_-^2 A_{NH}+ O(\epsilon_\pm^3),
\end{equation}
with $A_{NH} = c^2_{12} c^2_{23} s^2_{12} = {1\over 4} \sin^2 2 \theta_{\rm sol} \cos^2 \theta_{\rm atm} \sim 0.1$.
On the other hand, in the inverted hierarchy case, $\theta_{12} \sim \pi/2$. In the limit $\theta_{12} = \pi/2$ we get: 
\begin{equation}
\label{eq:pemumSSIH}
P_{e\mu}^{SS,IH} = 4 \sin^2 \left ( \frac{M^2 L}{4 E_\nu}\right ) \left (\epsilon_+-\epsilon_-\right )^2 A^{IH} + O(\epsilon_\pm^3),
\end{equation}
with $A_{IH} = c^2_{13} s^2_{13} s^2_{23} = {1\over 4} \sin^2 2 \theta_{\rm sol} \sin^2 \theta_{\rm atm}\sim 0.1$.  The inverted hierarchy  gets therefore an additional suppression with respect to the normal hierarchy case due to the partial cancellation between $\epsilon_+$ and $\epsilon_-$. It is easy to see that 
\begin{eqnarray}
(\epsilon^{IH}_+ - \epsilon^{IH}_-) \sim {|\Delta m^2_{sol}|\over |\Delta m^2_{atm}|} ~\epsilon_-^{NH}.
\end{eqnarray}
and, therefore, a suppression of the order of the solar to atmospheric mass ratio is to be expected in the inverted hierarchy case with respect to the normal hierarchy one. Unfortunately this accidental suppression for the IH is responsible for the fact that the leading-order result of eq.~(\ref{eq:pemumSSIH}) is not sufficiently precise and higher orders are relevant, but in any case the effect is too small. 
The numerical expressions for $P_{e\mu}$ for $L = 541$ m in the range of energies relevant for MiniBooNE are shown in Fig.~\ref{fig:MB} for the NH and IH, compared with the best fit result of \cite{Kopp:2011qd}. The strong suppression for the IH case is clearly seen.
The degenerate case cannot accommodate the LSND anomaly. 

A different situation is found in reactors, which actually set stronger constrains on $M^{SS}_{min}$. The reason is that the effects on disappearance are only linear in $\epsilon$ and not quadratic. In the same perturbative expansion as before we find:
\bea
P_{ee}^{SS}&=& 1-4\left[\epsilon_{+} s_{13}^2 +\epsilon_{-}c_{13}^2 s_{12}^2 \right]~\sin^2\frac{M^2L}{4E},
\nonumber\\
P_{\mu\mu}^{SS}&=& 1-4\left[\epsilon_{+}c_{13}^2 s_{23}^2+\epsilon_{-}(c_{12} c_{23} -s_{12}s_{13}s_{23})^2\right]~\sin^2\frac{M^2L}{4E}.
\eea
This is the origin of the well-known tension between the LSND appearance signal and the disappearance constraints, in the electron channel by the Bugey-3 experiment \cite{Declais:1994su} and in the muon channel by the CDHSW experiment \cite{Dydak:1983zq}. 

\begin{figure}[!ht]
\begin{center}
\includegraphics[width=15cm]{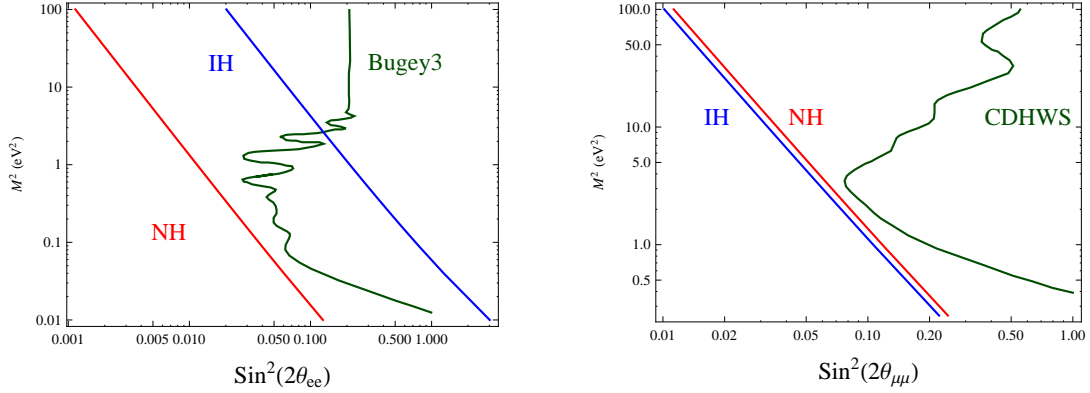} 
\vspace{0.5cm}
\end{center}
\caption{Exclusion region at $99\%$CL from the experiments Bugey-3  (left) and CDHSW  (right). The lines correspond to the expectation as a function of $M$ in the degenerate case. The intersection of the IH curve and the Bugey-3 exclusion region is at $M\sim 1.6$ eV. } 
\label{fig:disap}
\end{figure}

 In Fig.~\ref{fig:disap} we compare the exclusion plot of Bugey-3 and CDHSW with the prediction for the effective mixing angle and frequency in the disappearance channels in the degenerate case  (note that the two are related). We can see that there is a very small additional exclusion of the allowed region for the IH, which sets  
\be
M_{\rm min}^{SS} \sim 1.6 ~{\rm eV},
\ee
while there is no further constrain for the NH case. Note that these bounds would be relaxed if the new computation of the reactor fluxes is taken into account. 

\subsection{Other bounds}

It is clear that important constraints on the mini-seesaw region could also come from tritium $\beta$-decay \cite{Kraus:2004zw,pdg}, which is sensitive to the combination\footnote{It is assumed that all $m_i \ll \Delta E$, where $\Delta E$ is the energy resolution near the end-point. When non degenerate neutrinos are considered the analysis is more complicated \cite{Farzan:2001} and cannot be cast as an upper bound on the combination $m_e$.} 
\be
m_{e} = \sqrt{\sum_{i} |U_{ei}|^2 m_{\nu i}^2 } \leq 2~{\rm eV},
\label{eq:me}
\ee
and from cosmology. The amplitude of the neutrinoless double $\beta$-decay process on the other hand approximately vanishes for $M \ll {\mathcal O}(100)$ MeV and therefore cannot constrain further the mini-seesaw region (see  \cite{Blennow:2010th}).

Naively one would expect that such bounds will be in the form of a maximal value of $M^{SS}_{\rm max}$ in the mini-seesaw region.

Concerning tritium $\beta$-decay, at leading order in the $\epsilon$ expansion we get 
\be
\label{eq:mee}
m_{e} \simeq \sqrt{s_{13}^2 m^2_{D^+} + c_{13}^2 s_{12}^2 m_{D^-}^2 }.
\ee
 For the NH we therefore have 
\be
m_{e} \simeq \sqrt{s_{13}^2 \sqrt{\Delta m^2_{atm}} + c_{13}^2 s_{12}^2 \sqrt{\Delta m^2_{sol}} } \sqrt{M} \simeq 0.05~\sqrt{M({\rm eV})} {\rm eV}.
\ee
For the IH instead ($\theta_{12} \simeq \pi/2$ and $\theta_{13} \simeq \theta_{sol}$):
\be
m_{e} \simeq \sqrt{(s_{13}^2  + c_{13}^2 s_{12}^2) \sqrt{\Delta m^2_{atm}} } \sqrt{M} \simeq 0.22~\sqrt{M({\rm eV})} {\rm eV}.
\ee
Therefore, the present bound for the degenerate case gives $M^{SS}_{\rm max} \leq 82~ {\rm eV}$ (IH) and $M^{SS}_{\rm max} \leq 1.6 ~{\rm keV}$ (NH). These estimates are however too naive \cite{deGouvea:2006gz}, because for such large values of $M$, the heavy states are far from the end-point and the sensitivity to the heavy states cannot be cast into the form of their contribution to $m_e$ in eq.~(\ref{eq:me})\cite{Farzan:2001} .  In fact, a more carefull analysis along the lines of \cite{deGouvea:2006gz} shows that there is no bound on $M$ from the present measurements of tritium experiments, neither for the NH not for the IH.

More stringent bounds are expected from cosmology, that imposes stringent constraints on the number of thermalized relativistic species at the time 
of nucleosynthesis (BBN), and at recombination (CMB) \cite{Olive:1981ak}. Furthermore the large scale structure is already sensitive to neutrino masses in the eV range. Although a detailed calculation is required, both sets of measurements will very likely add new information on $M^{SS}_{\rm max}$ maybe closing the eV window for this model.

\section{Towards the non-degenerate case}
\label{sec:nodeg}

Although a systematic exploration of the parameter space in the non-degenerate case will be  studied elsewhere, in this section we want to point out the usefulness of  the parametrization introduced in sec.~\ref{sec:param}, which makes it easy to study analytically in the mini-seesaw region the oscillation probabilities in the same perturbative expansion introduced before. We consider  as an illustration the $P_{e\mu}^{SS}$ and $P_{ee}^{SS}$ probabilities to show  that the $n_R=2$ model in the non-degenerate case could explain the LSND/MB/reactor anomaly similarly to the solution found in \cite{Kopp:2011qd}.  

We consider the expansion  up to ${\mathcal O}(\epsilon^2 )$, where $\epsilon$ is any of the ratios of the light to heavy mass eigenstates
\be
\epsilon_{ij}\equiv m_i/M_j,
\ee
with $m_i$  the entries in eq.~(\ref{eq:msqrt}). These parameters reduce to $\epsilon_\pm$ in the degenerate limit. The results at leading non-trivial order are
\bea
P^{SS}_{e\mu}&=&4|\theta_{e4}|^2|\theta_{\mu4}|^2\sin^2\frac{M_1^2L}{4E}
+4|\theta_{e5}|^2|\theta_{\mu5}|^2 \sin^2\frac{M_2^2L}{4E}\nonumber\\
&+&8|\theta_{e4}\theta_{\mu4}\theta_{e5}\theta_{\mu5}|
\cos\left(\frac{(M_2^2-M_1^2)L}{4E}+\phi\right)\sin\frac{M_2^2L}{4E}\sin\frac{M_1^2L}{4E}+...\,,\label{Pnd}
\\
P^{SS}_{\alpha\alpha}&=&1-4|\theta_{\alpha4}|^2\sin^2\frac{M_1^2L}{4E}
-4|\theta_{\alpha5}|^2 \sin^2\frac{M_2^2L}{4E}
+ ...\,, 
\label{Pndaa}
\eea
where $\phi=\text{arg}\left(\theta_{e4}\theta_{\mu 4}^*\theta_{e5}^*\theta_{\mu 5}\right)$. The CP-conserving case corresponds to $\phi$ equal $\pi$ or $0$. We introduce the CP non conservation only through $\phi$ as a first step \footnote{A specific parametrization in terms of CP phases in the mixing matrix, will lead generically to phase dependence also in the $|\theta_{\alpha i}|$.}.  Using eq.~(\ref{eq:magic})
\be
\theta=\tilde{U}(\theta_{12},\theta_{13},\theta_{23}) m^{1/2}\tilde{W}(\theta_{45})^T M^{-1/2}_N,\,
\ee
 it is straightforward to obtain $\theta_{\alpha i}$ at the required order. 

The appearance probability is
\bea
\label{Pnd2}
P^{SS}_{e\mu}&=&4\,A_{44}\sin^2\frac{M_1^2L}{4E}
+4\,A_{55} \sin^2\frac{M_2^2L}{4E}\nonumber\\
&+&8\,\sqrt{A_{44}A_{54}}
\cos\left(\frac{(M_2^2-M_1^2)L}{4E}+\phi\right)\sin\frac{M_2^2L}{4E}\sin\frac{M_1^2L}{4E}\,,
\label{eq:pemundeg}
\eea
where, for NH and $\theta_{13}=0$:
\bea
\label{ANH}
A^{NH}_{44}&=& \epsilon_{21} s_{12}^2s_{45}^2 \left(\epsilon_{31}^{1/2}c_{45}s_{23}+
\epsilon_{21}^{1/2}c_{12}c_{23}s_{45}\right)^2,
\nonumber\\
A^{NH}_{55}&=& \epsilon_{22} s_{12}^2 c_{45}^2  \left(\epsilon_{22}^{1/2}c_{12}c_{23}c_{45}
-\epsilon_{32}^{1/2}s_{23}s_{45}\right)^2,
\eea
and for IH and $\theta_{12}=\pi/2$:
\bea
\label{AIH}
A^{IH}_{44}&=& s_{23}^2 \left(\epsilon_{31}^{1/2}c_{45}s_{13}+
\epsilon_{21}^{1/2}c_{13}s_{45}\right)^2 \left(\epsilon_{31}^{1/2}c_{13}c_{45}-
\epsilon_{21}^{1/2}s_{13}s_{45}\right)^2,
\nonumber\\
A^{IH}_{55}&=& s_{23}^2  \left(\epsilon_{22}^{1/2}c_{45}s_{13}+
\epsilon_{32}^{1/2}c_{13}s_{45}\right)^2 \left(\epsilon_{22}^{1/2}c_{13}c_{45}-
\epsilon_{32}^{1/2}s_{13}s_{45}\right)^2,
\eea
The electron disappearance probabilities are
\bea
\label{Pndee}
\left.P_{ee}^{SS}\right|_{NH}&=&1-4\sin^2 \theta_{12}\left(\epsilon_{21}s_{45}^2\sin^2\frac{M_1^2L}{4E}
+\epsilon_{22}c_{45}^2\sin^2\frac{M_2^2L}{4E}\right),
\\
\left.P_{ee}^{SS}\right|_{IH}&=&1-4\left[\left(\epsilon_{31}^{1/2}c_{45}s_{13}+\epsilon_{21}^{1/2}s_{45}c_{13}\right)^2
\sin^2\frac{M_1^2L}{4E}
+\left(\epsilon_{22}^{1/2}c_{45}c_{13}-\epsilon_{32}^{1/2}s_{45}s_{13}\right)^2\sin^2\frac{M_2^2L}{4E}\right].\nonumber
\eea
It is  easy to check that these results coincide in the degenerate limit with those in the previous section, as they should. 

\begin{figure}[!ht]
\begin{center}
\includegraphics[width=15cm]{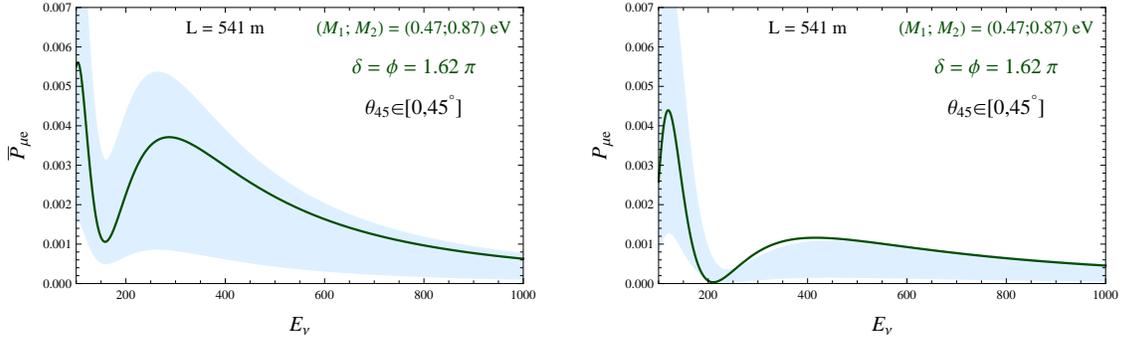}
\vspace{0.5cm}
\end{center}
\caption{Comparison of the energy dependence of the appearance probabilities $P_{\mu e}$ for neutrinos (right) and antineutrinos (left) at the MiniBoone baseline (L=541m)  for the solution KMS \cite{Kopp:2011qd} (solid line) and the non-degenerate case  (IH) fixing $M_1$ and $M_2$ to the best fit values of \cite{Kopp:2011qd}. The bands correspond to varying the free parameter $\theta_{45}$. }
\label{fig:nondegemu}
\end{figure}

\begin{figure}[!ht]
\begin{center}
\includegraphics[width=9cm]{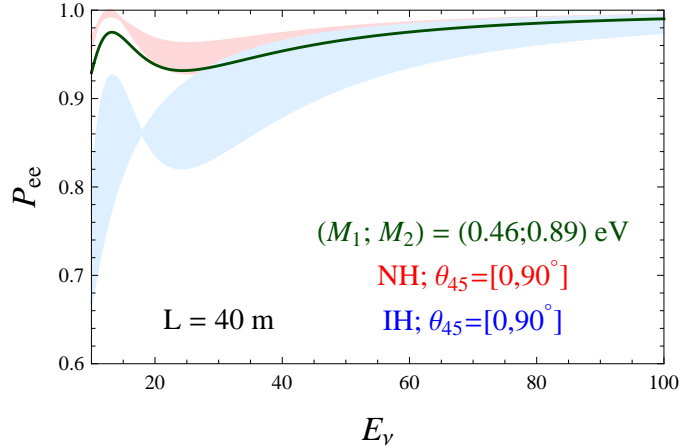}
\vspace{0.5cm}
\end{center}
\caption{Comparison of the energy dependence of the electron disappearance probabilities at the Bugey-3 baseline (L=40 m)  for the solution of \cite{Kopp:2011qd} (solid line) and the non-degenerate case fixing $M_1$ and $M_2$ to the best fit values of \cite{Kopp:2011qd}. The bands correspond to varying the free parameter $\theta_{45}$. }
\label{fig:nondegee}
\end{figure}

 Setting the two heavy states to the masses indicated in \cite{Kopp:2011qd}, it turns out that the angles are in the right ballpark for the IH. For the NH on the other hand, they are too small. In Fig.~\ref{fig:nondegemu} we compare the probability for the best fit of \cite{Kopp:2011qd}. We find that for $\theta_{45}\sim 20^\circ$ and $\phi=1.62 \pi$ the results for the appearance probabilities are quite similar and therefore the model can also reproduce the LSND/MiniBoone anomaly with a similar level of agreement as \cite{Kopp:2011qd}. The signal in disappearance shown in Fig.~\ref{fig:nondegee} is slightly larger so probably there is more tension between appearance and disappearance.

Finally and in the same expansion we obtain for the tritium $\beta$ decay effective mass of eq.~(\ref{eq:me}), at the leading order 
\bea
m^2_{e}&=&M_1\left(m_3^{1/2}c_{45}s_{13} +m_2^{1/2}c_{13}s_{12}s_{45}\right)^2 + 
 M_2\left(m_2^{1/2}\,c_{13}c_{45}s_{12} - m_3^{1/2}s_{13}s_{45}\right)^2 .
\eea
Fixing the free parameter $\theta_{45}$ to zero, a simpler expression which is the same as in the degenerate case, eq.~(\ref{eq:mee}), is obtained. For the KMS best fit choice of $M_1$ and $M_2$ we obtain $m_{e} \sim 0.05$ eV for NH and $m_{e} \sim 0.2$ eV for IH.

\section{Quasi-Dirac neutrinos in the sun}
\label{sec:solar}

In this section, we do an analytical study of the flavour transitions of quasi-Dirac neutrinos (solutions of Types I and II) in the sun in order to understand the very stringent limits on the value of $M$ set by solar data. A similar study has been discussed in \cite{deGouvea:2009fp}, see also \cite{deHolanda:2010am}. Older studies of pseudo-Dirac neutrinos in the sun were done in \cite{Nir:2000xn,GonzalezGarcia:2000ve}. 
 
 In order to understand the reason for the sensitivity of the sun to tiny splittings in the standard three Dirac neutrino scenario, we consider the NH case with  $\theta_{13}=0$, which is a good approximation. The IH works similarly, after the appropriate mapping in the angles (in particular, $\theta_{13}=0$ is equivalent to $\theta_{12}=\pi/2$), so in order to simplify the discussion we just consider the NH case in the intermediate results and indicate the differences in the final results for the IH.
 
 Let us first consider the case $M=0$, where the standard scenario must be recovered.  
 The eigenvalues in matter are:
\begin{eqnarray}
   \mu^{(0)} _0 &=& {1 \over 2} \left(A+ m_{D^-}^2 - \sqrt{A^2 + m_{D^-}^4 - 2 A m_{D^-}^2 \cos 2 \theta_{12}}\right)\nonumber\\\mu^{(0)}_1&=& {B \over 2}  + m_{D^-}^2\nonumber\\
    \mu^{(0)}_2 &=&{1 \over 2} \left(A+ m_{D^-}^2 + \sqrt{A^2 + m_{D^-}^4 - 2 A m_{D^-}^2 \cos 2 \theta_{12}}\right)\nonumber\\
     \mu^{(0)}_3 &=& m_{D^+}^2\nonumber\\
  \mu^{(0)}_4 &=& {B \over 2} +  m_{D^+}^2\nonumber\\
\end{eqnarray}
where $A= 2 \sqrt{2} G_F N_e E_\nu$ and $B= 2 \sqrt{2}  G_F N_n E_\nu$.  Since we are interested in a rough estimate, we will neglect $N_n$ in front of  $N_e$, that is  $B$ in front of $A$, since in the sun this is not too bad an approximation. 
The corresponding eigenvectors, $v_i^{(0)}(A)$, fall in three invariant subspaces
: the $1$, the $4$ and $023$. The standard 3$\nu$ result corresponds to the $023$ sector where the usual two-family MSW  is found in the sector 02. The MSW resonance occurs at $A_{MSW}= m_{D^-}^2 \cos 2 \theta_{12}$. 
However, the eigenvalue $\mu^{(0)}_1$ becomes exactly degenerate
with $\mu^{(0)}_2$ at $A=0$.
For $M=0$, there is an exact level crossing because the levels do not mix. When $M$ is non-zero, no matter how small, there is mixing and therefore an additional MSW effect in the sector 12 takes place, where the mixing is controlled by $M$, as we will see.
 
 It is found that only the states $v_i^{(0)}$ with $i=0,2$ have an electron component, and therefore the electron flavour state produced at the center of the sun is a mixture of  both $v_i^{(0)}(A_0)$,  where  $A_0$ corresponds to the value of $A$  close to the center of the sun.
In the adiabatic approximation, therefore, the probability $\nu_e \rightarrow \nu_e$ (ignoring possible effects due to Earth matter) is:
\begin{eqnarray}
P_{ee} &=& \left[(v^{(0)}_0(A_0))_e\right]^2 \left[(v^{(0)}_0(0))_e\right]^2+ \left[(v^{(0)}_2(A_0))_e\right]^2 \left[(v^{(0)}_2(0))_e\right]^2 \nonumber\\
&=&{1\over 2} +
{\cos 2 \theta_{12} \over 2} { m_{D^-}^2 \cos 2 \theta_{12} -A_0 \over  \sqrt{A_0^2 + m_{D^-}^4 - 2 A_0 m_{D^-}^2 \cos 2 \theta_{12}}}.
\end{eqnarray}
We recover the expected Dirac result:
\begin{eqnarray}
P_{ee} \rightarrow \left\{ \begin{array}{ll}  \sin^2 \theta_{12}  & A_0 \gg A_{MSW}, \\
 1- {1\over 2} \sin^2 2 \theta_{12}  & A_0 \ll A_{MSW}. 
 \end{array}\right.
\end{eqnarray}
The result for the IH case is equivalent after exchanging $m_{D^-}^2 \rightarrow m_{D^+}^2 - m_{D^-}^2$ and $\theta_{12} \rightarrow \theta_{13}$. The relevant mass scale is therefore in both cases $\Delta m^2_{\rm sol}$.

Let us now consider a value of $M \ll m_{D^-} <  m_{D^+}$. We can  do perturbation theory to include the effects of $M$ for any value of $A$, but for $A \rightarrow 0$, the pairs of states $12$ and $34$ are degenerate,  and therefore standard perturbation theory breaks down. Non-degenerate perturbation theory is necessary here, which amounts to  diagonalizing the perturbation matrix exactly in the corresponding $2\times2$ subspaces, treating the rest in standard perturbation theory. The resulting eigenvalues and eigenvectors are denoted by $\mu_i^{(M)}$ and $v_i^{(M)}$. If we neglect all effects of ${\mathcal O}(M/m_{D^\pm})$, we have that, for large $A$,
the eigenstates go smoothly to those at $M= 0$:
\be
\lim_{M\rightarrow 0}~ v^{(M)}_i(A) = v^{(0)}_i(A).
\ee
However, for small $A$, more concretely for $A \ll M m_{D^\pm}$, the $M \rightarrow 0$ limit does not coincide with the $M=0$ case:
\begin{eqnarray}
\lim_{M\rightarrow 0} \left(v_1^{(M)}(0)\right)_e &=& -{\sin \theta_{12} \over \sqrt{2}}, \nonumber \\
\lim_{M\rightarrow 0} \left(v_2^{(M)}(0)\right)_e &=& -{\sin \theta_{12} \over \sqrt{2}}, 
\label{eq:disc}
\end{eqnarray}
while $(v_1^{(0)}(0))_e =0$.

In the adiabatic limit, we find 
\begin{eqnarray}
\lim_{M \rightarrow 0} P_{ee} &=&\lim_{M\rightarrow 0} \left[(v^{(M)}_0(A_0))_e\right]^2 \left[(v^{(M)}_0(0))_e\right]^2+ \left[(v^{(M)}_2(A_0))_e\right]^2 \left[(v^{(M)}_2(0))_e\right]^2\nonumber\\
&=& \left({\sin^2 \theta_{12}\over 4} + { \cos^2 \theta_{12}\over 2} \right) +  \left({\sin^2 \theta_{12}\over 4} - { \cos^2 \theta_{12}\over 2} \right) {A_0-m_{D^-}^2 \cos2\theta_{12}\over \sqrt{A_0^2 + m_{D^-}^4 - 2 A_0 m_{D^-}^2 \cos 2 \theta_{12} }}.\nonumber
\end{eqnarray}
Therefore
\begin{eqnarray}
\left. P_{ee} \right|_{NH} \rightarrow \left\{ \begin{array}{ll}  {\sin^2 \theta_{12}\over 2}  & A_0 \gg A_{MSW}, \\
 \cos^4 \theta_{12} +{ \sin^4 \theta_{12}\over 2} & A_0 \ll A_{MSW}. 
 \end{array}\right.
\end{eqnarray}
  Note that for energies above  the MSW resonance $P_{ee}$ is  1/2 of the Dirac result, so, to the extent that the adiabatic approximation is valid, the result for $M\neq 0$ is physically very different to the Dirac limit. 
  
  A similar analysis in the IH case is a bit more complicated, because in this case the problem involves the two pairs of degenerate states and not just one pair as before. The final result is
   \begin{eqnarray}
\left. P_{ee}\right|_{IH} \rightarrow \left\{ \begin{array}{ll}  {\sin^2 \theta_{13}\over 2}  & A_0 \gg A_{MSW}, \\
 {\cos^4 \theta_{13}\over 2} +{ \sin^4 \theta_{13}\over 2} & A_0  \ll A_{MSW}. 
 \end{array}\right. ,
\end{eqnarray}
where the MSW condition is the usual one with the change $m_{D^-}^2 \rightarrow m_{D^+}^2 - m_{D^-}^2$. Note that in our parametrization, $\theta_{13}$ is the solar angle for the IH case. In this case, the result is half of the Dirac result  both above and below MSW.

Obviously we expect  the limit $M\rightarrow 0$ to be smooth and this means that adiabaticity must break down for small enough $M$. We show that this is indeed the case. 
 \subsection{Adiabaticity limit}

Adiabaticity is lost in the propagation inside the sun when 
\begin{eqnarray}
{|\mu^{(M)}_1 - \mu^{(M)}_2| \over 2 E_\nu} <  \left|v^{(M)}_1 \cdot {d \over d A} v^{(M)}_2  {d A \over d r}\right|,
\label{eq:nonadi}
\end{eqnarray}
where $r$ is the radial distance. 
For $m_{D^\pm}$ large compared to the other scales we have
\begin{eqnarray}
 |\mu^{(M)}_1 - \mu^{(M)}_2|  &\simeq& \sqrt{4 M^2 m^2_{D^-} + A^2 \cos^4 \theta_{12}} \nonumber\\
 \left|v^{(M)}_1 \cdot {d \over d A} v^{(M)}_2  \right| &\simeq& {M m_{D^-} \cos^2  \theta_{12} \over 4 M^2 m^2_{D^-} + A^2 \cos^4 \theta_{12}}
\label{eq:eiga0}
\end{eqnarray}
The variation of $A$ can be approximated by 
\begin{eqnarray}
{d A \over d r} \simeq - \alpha A/R_\odot,
\end{eqnarray}
where $\alpha \simeq 10-15$ and $R_\odot$ is the solar radius. 

The right term in eq.~(\ref{eq:nonadi}) is maximal at the point in the evolution 
where 
\begin{eqnarray}
A^2 \cos^4 \theta_{12} \sim 4 M^2 m_{D^-}^2. \; \; \
\label{eq:maxna}
\end{eqnarray}
At this point the non-adiabaticity condition reads
\begin{eqnarray}
M \leq {E_\nu \alpha \over 4 \sqrt{2} R_\odot m_{D^-}}.
\label{eq:adi}
\end{eqnarray}
 The result for the IH is the same with the change $ \theta_{12} \rightarrow  \theta_{13}$, but since $m_{D^-} \sim \sqrt{\Delta m^2_{\rm sol}}$ for NH and $m_{D^-} \sim \sqrt{\Delta m^2_{\rm atm}}$ for the IH, the value of $M$ for which the adiabaticity limit is reached is lower for IH than for NH, by the ratio of solar to atmospheric mass splittings. The rough estimates are 
\begin{eqnarray}
M({\rm eV}) < \left\{\begin{array}{lll}    10^{-7} \times E_\nu({\rm MeV})    & & {\rm NH},\\
   2 \times 10^{-8} \times E_\nu({\rm MeV})    &  & {\rm IH}. \end{array} \right.
\end{eqnarray}

\subsection{Dirac Limit}

Now that we have established that adiabaticity is lost, we must still prove that the Dirac limit is reached for such small $M$. An easy way to see how this happens  is to consider   the sudden approximation. The evolution of the eigenstates  shows an abrupt change in the 12 sector at the condition of eq.~(\ref{eq:maxna}). For larger $A$ the eigenstates evolve smoothly and are close to $v_i^{(0)}(A)$, but at this threshold, the eigenstates 1 and 2 change abruptly to be similar to those in $v_i^{(M)}(0)$.  The sudden approximation assumes that the physical states do not change abruptly and therefore a physical state that is an eigenstate as it approaches the transition for $A > A_{th}$ does not remain in an eigenstate for $A < A_{th}$. Oscillations therefore occur.

At the center of the sun the electron neutrino is a combination of $v^{(M)}_0(A_0)\simeq v^{(0)}_0(A_0)$ and $v^{(M)}_2(A_0)\simeq v^{(0)}_2(A_0)$ eigenstates. The two very fast decohere and the two components evolve adiabatically until the transition where 
\begin{eqnarray}
v^{(M)}_2(A^+_{th})  \simeq  \left(  {1 \over \sqrt{2}}v^{(M)}_1(A^-_{th}) + {1\over \sqrt{2}} v^{(M)}_2(A^-_{th}) \right) \simeq  \left(  {1 \over \sqrt{2}}v^{(M)}_1(0) + {1\over \sqrt{2}} v^{(M)}_2(0) \right) ,
\end{eqnarray} 
where we have neglected effects of ${\mathcal O}(A_{th}/m^2_{D^\pm}, M/m_{D^\pm})$. The evolution of the 0 state is adiabatic throughout (this is different for the IH case). 

The time evolution of the state after the transition point at $t_0$ is approximately
\begin{eqnarray}
 {1\over \sqrt{2}} v^{(M)}_1(0) e^{-i {\mu^{(M)}_1(0)\over 2 E_\nu} (t-t_0) } +{1\over \sqrt{2}} v^{(M)}_2(0) e^{-i{ \mu^{(M)}_2(0) \over 2 E_\nu}(t-t_0) } .  
\end{eqnarray} 
Using eq.~(\ref{eq:disc}), the $P_{ee}$ probability in the sudden approximation is found to be:
\begin{eqnarray}
\left.P_{ee}\right|_{NH} =\left\{\begin{array}{ll} \sin^2 \theta_{12} \cos^2\left[ {M m_{D^-}\over 2 E_\nu} \Delta t \right] & A_0 \gg A_{MSW} \\
 \cos^4 \theta_{12} + \sin^4 \theta_{12} \cos^2\left[ {M m_{D^-}\over 2 E_\nu} \Delta t\right] & A_0 < A_{MSW}. \\
\end{array}\right.
\end{eqnarray}
To result for the IH  is instead
\begin{eqnarray}
\left.P_{ee}\right|_{IH} =\left\{\begin{array}{ll} \sin^2 \theta_{12} \cos^2\left[ {M m_{D^-}\over 2 E_\nu} \Delta t \right] & A_0 \gg A_{MSW} \\
 \cos^4 \theta_{12}  \cos^2\left[ {M m_{D^+}\over 2 E_\nu} \Delta t\right]+ \sin^4 \theta_{12} \cos^2\left[ {M m_{D^-}\over 2 E_\nu} \Delta t\right] & A_0 < A_{MSW}. \\
\end{array}\right.
\end{eqnarray}

\begin{figure}[!ht]
\begin{center}
\includegraphics[width=5.cm]{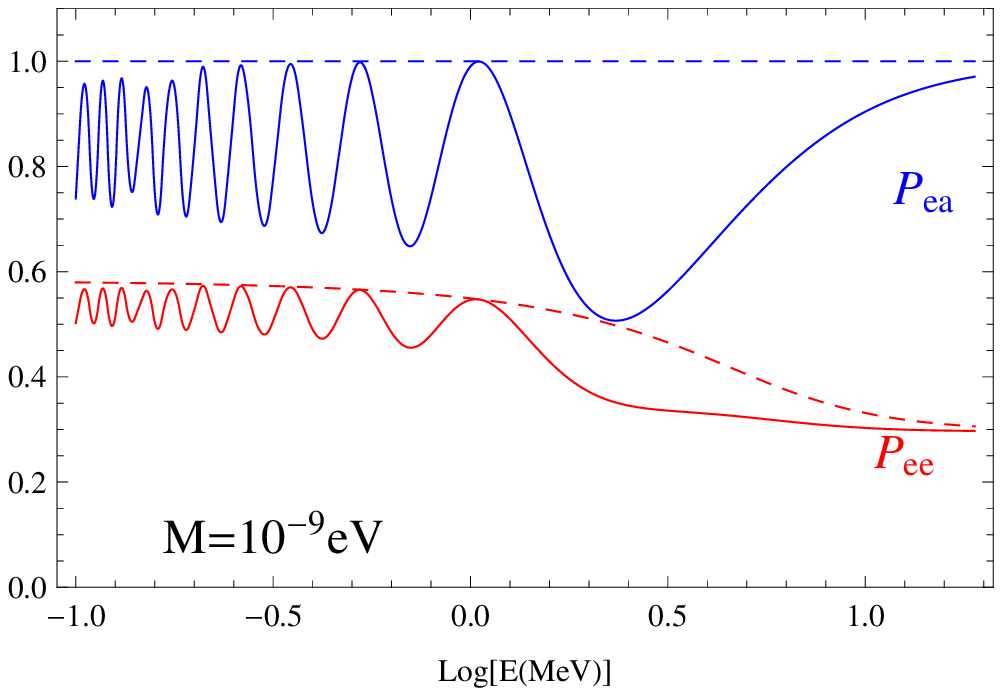}\includegraphics[width=5.cm]{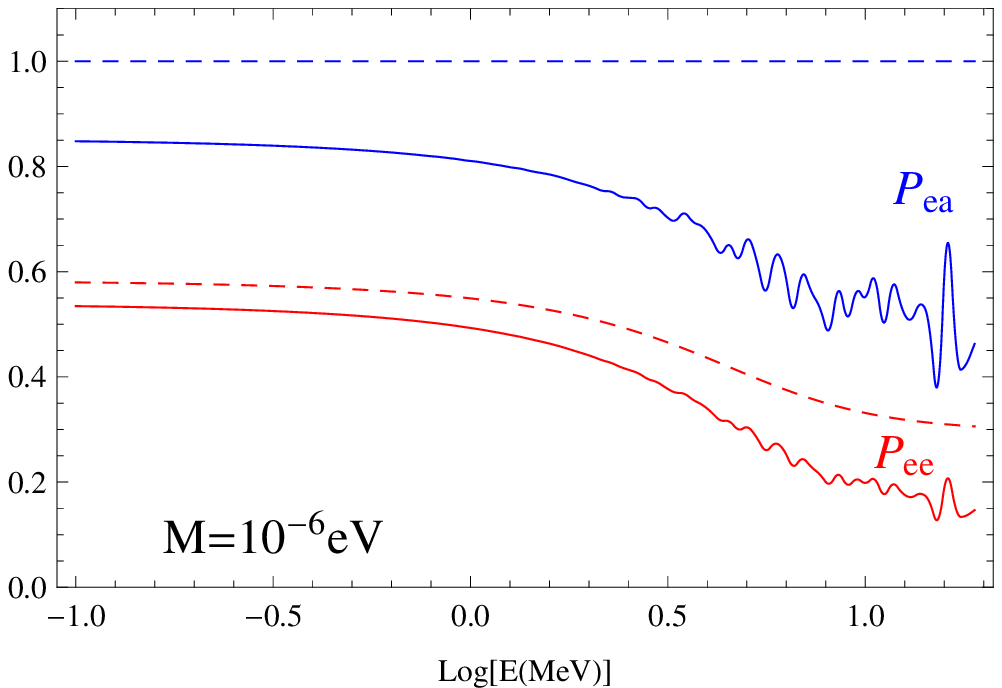}\includegraphics[width=5.cm]{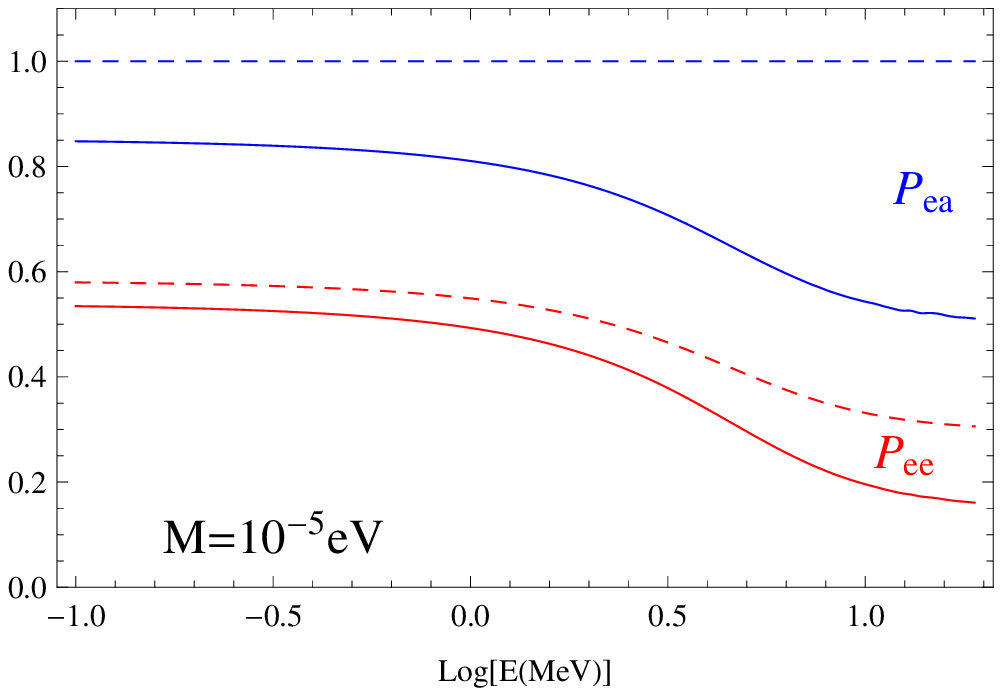}
\vspace{0.5cm}
\end{center}
\caption{ For the NH, $P_{ee}$ (red), $P_{ea}=P_{ee}+P_{e\mu}+P_{e\tau}$ (blue)  at day time  as a function of neutrino energy for solar neutrinos (B) in the quasi-Dirac region for three values of $M=10^{-9},10^{-6},10^{-5}$eV and normal hierarchy. The dashed curves correspond to the standard 3$\nu$ solution (Dirac limit) near the best fit and the solid lines are the exact results in the quasi-Dirac for the same values of the parameters. The production point has been averaged out. }
\label{fig:probsnh}
\end{figure}

\begin{figure}[!ht]
\begin{center}
\includegraphics[width=5.cm]{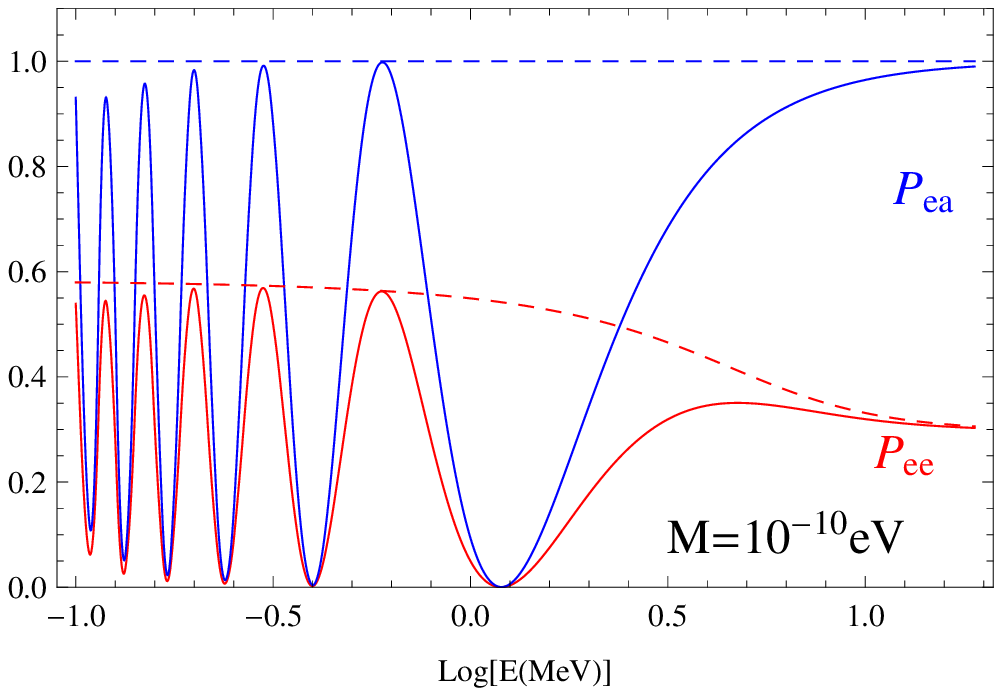}\includegraphics[width=5.cm]{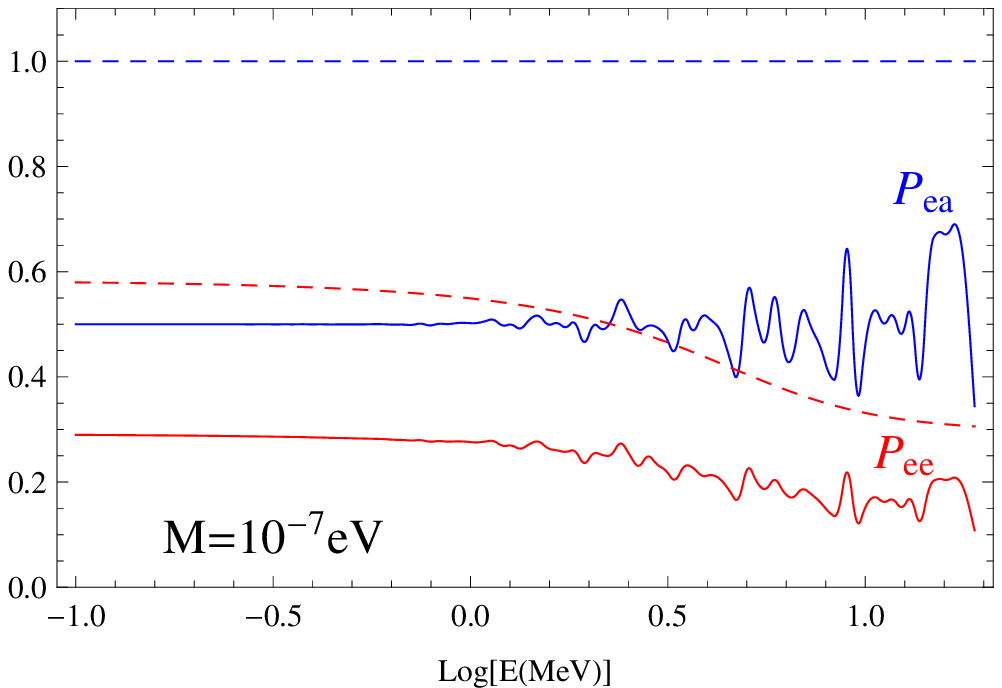}\includegraphics[width=5.cm]{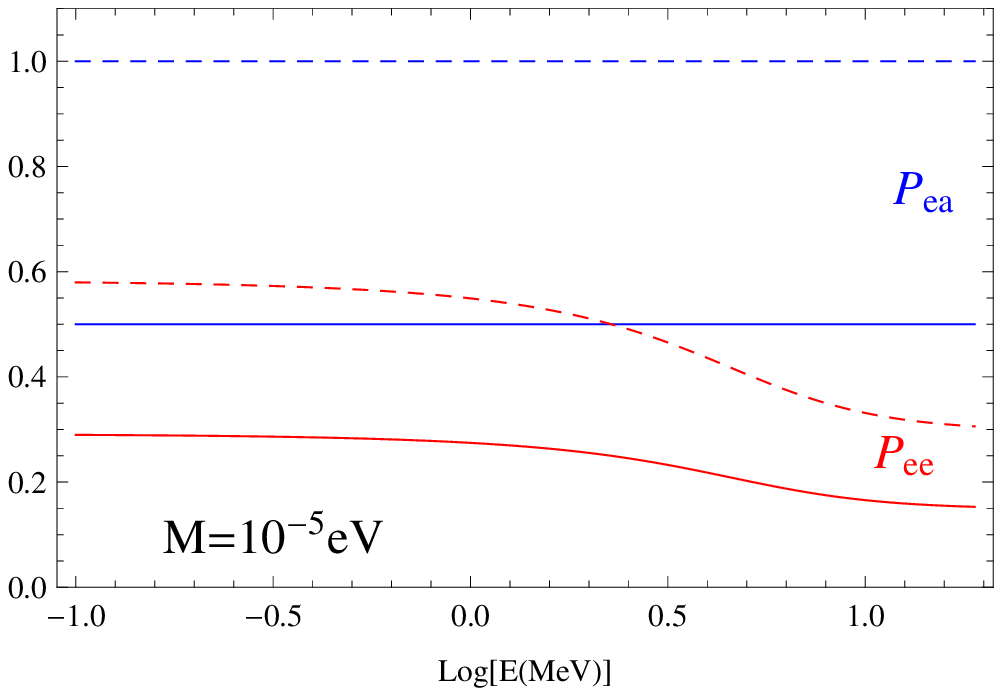}
\vspace{0.5cm}
\end{center}
\caption{ The same as Fig.~14 for the IH.}
\label{fig:probsih}
\end{figure}

When the oscillatory terms are approximately one, the  Dirac limit is reached, while the regime of fast oscillations reduces to the 
 adiabatic result. In order to recover Dirac we need two conditions:
1) non adiabaticity and 2) long enough baseline for unaveraged vacuum oscillations with the splitting $M m_{D^-}$. At the adiabaticity limit (eq.~(\ref{eq:adi})), the oscillation length  is roughly the solar radius for all energies, and therefore the vacuum oscillations are averaged out on Earth. Smaller values of $M$ are necessary to increase the oscillation length. 
These features are precisely found in the exact $P_{ee}$ and $P_{ea}$ shown in Fig.~\ref{fig:probsnh} for the NH and in Fig.~\ref{fig:probsih} for the IH. For $M\sim 10^{-5}$eV, the adiabatic result of eq.~(\ref{eq:adi}) is found for all the energies shown. For $M \sim10^{-6}$eV the  vacuum oscillations are seen for larger energies while there are averaged oscillations still for the lower energies. For smaller values
of $M$, the vacuum oscillations are seen at all energies, and for $M\leq M_{max}^{QD}$ the Dirac limit is obtained. 

\vspace{0.5cm}

\vspace{0.5cm}

\section{Conclusions and Outlook}

Probably the simplest explanation of neutrino masses involves the addition of singlet fermions to the SM. Such is the case in very different  models ranging from Dirac neutrinos, to type I seesaws, inverse seesaw, etc. All these possibilities, that have very different phenomenological implications for flavour physics, correspond to different numbers of extra species and/or different global symmetries. It is important to study the present constraints on models with singlet fermions in increasing order of complexity, where complexity is measured by the number of extra degrees of freedom (Weyl fermions), $n_R$, and not by the physical spectrum, because the latter can depend also on the global symmetries imposed. With this perspective in mind, we have considered in the present work the two simplest possibilities, that of one or two extra Weyl fermions, $n_R =1,2$. The first case has sufficient free parameters (two masses and two mixing angles) to fit {\it in principle} the two confirmed oscillations (solar and atmospheric), but a complete scan of the parameter space shows that it cannot fit all the available data from long-baseline reactor and accelerator experiments (LBL). The case $n_R=2$ is equivalent to the standard 
$3\nu$-mixing scenario (with one massless neutrino) in two limiting cases: the Dirac limit (vanishing Majorana masses for the extra fields)
and the seesaw limit (large Majorana masses for the extra fields), and as such it does provide a good fit to the data in both limits. 
What happens in between these limiting cases is much more complicated. We have thoroughly studied the allowed parameter space in this model by requiring that at least one mass splitting corresponds to the solar one and another to the atmospheric one, and considered the simplifying assumption of degenerate Majorana masses, $M$.
 Even though many exotic solutions for intermediate values of $M$ can fit very well reactor and accelerator data, they are shown to fail to explain  solar or atmospheric neutrino data. We exclude therefore all regions except the quasi-Dirac, $M \lesssim M^{QD}_{max} = 10^{-9} (10^{-10})$ eV, and the mini-seesaw for $M \gtrsim M^{SS}_{min}= 0.6 (1.6)$ eV for the NH(IH) respectively. The upper bound, $M^{QD}_{max}$, is essentially constrained by solar data alone, while the lower bound $M^{SS}_{min}$ is set by LBL and reactor disappearance data.  A relevant upper bound to the mini-seesaw region could clearly be obtained from cosmology.
  
 We have discussed in detail the very stringent constraint imposed by solar neutrino experiments on quasi-Dirac neutrinos.
  
  The relevance of the mini-seesaw solutions to explain some of the unsolved neutrino anomalies: reactors, LSND,  MiniBoone is discussed and discarded in the degenerate case. Such anomalies cannot be explained in the context of the degenerate $n_R=2$ model within the parameter space allowed by other neutrino oscillation experiments. 
    However, we have argued that in the non-degenerate case, it is possible to obtain the pattern favoured by \cite{Kopp:2011qd}, although the same tension between appearance and disappearance still remains. The parametrization of Casas-Ibarra \cite{Casas:2001sr} in the mini-seesaw region of the general model has been used to derive accurate approximations of the oscillation probabilities in this regime.
        
    The detailed
    constraints for the non degenerate case including CP violation will be considered in future work.
    We have not explored  the restricted parameter space implied by a possible approximate lepton number symmetry,  that can induce technically natural hierarchies in the spectrum. Such approximate symmetry  could imply  cancellations that could have been missed in our scan, and need to be searched for more carefully.  Such possibility, as well as the model with increased level of complexity implied by an additional Weyl fermion, $n_R=3$, are interesting avenues to be explored. 

\acknowledgments

We thank  M. Blennow, E. Fern\'andez-Martinez, O. Mena, S. Pascoli and C.~Pe\~na-Garay for useful discussions. 
This work was partially supported by the Spanish Ministry for Education and Science projects  FPA2007-60323, FPA2009-09017; the Consolider-Ingenio CUP (CSD2008-00037) and CPAN (CSC2007-00042); the Generalitat Valenciana (PROMETEO/2009/116); the Comunidad Aut\'onoma de Madrid (HEPHACOS P-ESP-00346 and HEPHACOS S2009/ESP-1473);  the European projects EURONU (CE212372),  EuCARD (European Coordination for  
Accelerator Research and Development, Grant Agreement number 227579) and LAGUNA (Project Number 212343).
\bibliographystyle{h-elsevier}
\bibliography{biblio}

\end{document}